\documentclass{aastex}
\usepackage{emulateapj5}

\usepackage{epsfig}

\newcommand{\kms}{km~s$^{-1}$}
\newcommand{\lya}{Lyman~$\alpha$}
\newcommand{\lyb}{Lyman~$\beta$}
\newcommand{\lyg}{Lyman~$\gamma$}

\newcommand{\chid}{$\chi^2$}

\newcommand{\cl}{$95.5\%$ c.l.}
\newcommand{\ndi}{$N$(D~{\sc i})}
\newcommand{\noi}{$N$(O~{\sc i})}
\newcommand{\nni}{$N$(N~{\sc i})}
\newcommand{\nhi}{$N$(H~{\sc i})}

\begin{document}

\title{Deuterium abundance toward G191-B2B: Results from the Far Ultraviolet
Spectroscopic Explorer ({\it FUSE}) Mission
\footnote{This work is based on data obtained for the Guaranteed Time
Team by the NASA-CNES-CSA {\it FUSE} mission operated by the Johns
Hopkins University.}  }

\author{M.~Lemoine,
A.~Vidal--Madjar,
G.~H\'ebrard, 
J.-M.~D\'esert, 
R.~Ferlet, 
A.~Lecavelier des Etangs}
\affil{Institut d'Astrophysique de Paris, CNRS, 98 bis bld Arago, 
F-75014 Paris, France}
\author{J.C.~Howk,
M.~Andr\'e,
W.P.~Blair,
S.D.~Friedman,
J.W.~Kruk,
S.~Lacour,
H.W.~Moos,
K.~Sembach,
}
\affil{Department of Physics and Astronomy, Johns Hopkins University, 
Baltimore, MD 21218, USA}
\author{P.~Chayer}
\affil{UVIC/Department of Physics and Astronomy, 
Department of Physics and Astronomy, Johns Hopkins University, 
Baltimore, MD 21218, USA}
\author{E.B.~Jenkins}
\affil{Princeton University Observatory, Princeton, 
NJ 08544, USA}
\author{D.~Koester}
\affil{Institut f\"ur Theoretische Physik und Astrophysik, Universit\"at Kiel,
D-24098 Kiel, Germany}
\author{J.L.~Linsky, B.E.~Wood}
\affil{JILA, University of Colorado and NIST,
Boulder, CO 80309-0440, USA}
\author{W.R.~Oegerle, G.~Sonneborn}
\affil{Laboratory for Astronomy and Solar Physics, NASA/GSFC, Code 681, 
Greenbelt, MD 20771, USA}
\and
\author{D.G.~York}
\affil{Dept. of Astronomy \& Astrophysics, Univerity of Chicago, Chicago,
IL 60637, USA}

\begin{abstract}

High-resolution spectra of the hot white dwarf G191-B2B, covering the
wavelength region 905--1187\AA\/, were obtained with the Far
Ultraviolet Spectroscopic Explorer ({\it FUSE}). This data was used in
conjunction with existing high-resolution Hubble Space Telescope STIS
observations to evaluate the total H~{\sc i}, D~{\sc i}, O~{\sc i} and
N~{\sc i} column densities along the line of sight. Previous
determinations of $N$(D~{\sc i}) based upon GHRS and STIS observations
were controversial due to the saturated strength of the D~{\sc i}
Lyman~$\alpha$ line. In the present analysis the column density of
D~{\sc i} has been measured using only the unsaturated Lyman~$\beta$
and Lyman~$\gamma$ lines observed by {\it FUSE}. A careful inspection
of possible systematic uncertainties tied to the modeling of the
stellar continuum or to the uncertainties in the {\it FUSE}
instrumental characteristics has been performed. The column densities
derived are: $\log N($D~{\sc i}$)=13.40\pm0.07$, $\log N($O~{\sc
i}$)=14.86\pm0.07$, and $\log N($N~{\sc i}$)=13.87\pm 0.07$ quoted with
$2\sigma$ uncertainties.

  The measurement of the H~{\sc i} column density by profile fitting of
the Lyman~$\alpha$ line has been found to be unsecure. If additional
weak hot interstellar components are added to the three detected
clouds along the line of sight, the H~{\sc i} column density can be
reduced quite significantly, even though the signal-to-noise ratio and
spectral resolution at Lyman~$\alpha$ are excellent.  The new estimate
of \nhi\ toward G191-B2B reads: $\log N($H~{\sc i}$)=18.18\pm0.18$
($2\sigma$ uncertainty), so that the average (D/H) ratio on the line
of sight is: (D/H)$=1.66^{+0.9}_{-0.6}\times10^{-5}$ ($2\sigma$ uncertainty).
\end{abstract}


\section{Introduction}

Deuterium is thought to be produced in significant amount only during
primordial Big Bang nucleosynthesis (BBN), and to be thoroughly
destroyed in stellar interiors.  Deuterium is thus a key element in
cosmology and in Galactic chemical evolution (see e.g. Audouze \&
Tinsley 1976; Gautier \& Owen 1983; Vidal--Madjar \& Gry 1984;
Boesgaard \& Steigman 1985; Olive {\it et~al.} 1990; Pagel 1992;
Vangioni-Flam \& Cass\'e 1994; Vangioni-Flam {\it et~al.}  1995;
Prantzos 1996; Scully {\it et~al.} 1997; Cass\'e \& Vangioni-Flam
1998; Tosi {\it et al.} 1998).  Indeed, its primordial abundance is
the best tracer of the baryonic density parameter of the Universe,
$\Omega_B$, and the decrease in its abundance during galactic
evolution traces the cosmic star formation rate at various epochs.

The first (indirect) measurement of the deuterium abundance was
carried out using $^3$He in the solar wind, giving the presolar value
D/H$\simeq2.5\pm1.0\times10^{-5}$ (Geiss \& Reeves 1972). The first
measurements of the D/H ratio in the interstellar medium (ISM) were
reported shortly thereafter by Rogerson \& York (1973), and their
value (D/H)$\simeq1.4\pm0.2\times10^{-5}$ has remained a landmark
average value for the interstellar D/H ratio. Finally direct
measurements of the primordial (D/H) ratio in low-metallicity material
at high redshift have been successfully carried out these past few
years (e.g., Burles 2001 for a review, and references therein). The
values derived cluster around (D/H)$\sim 3\times10^{-5}$ although with
significant dispersion, which may or may not be real. Quite similarly
the measurements of the (D/H) ratio in the Galactic ISM towards hot
stars with the {\it Copernicus} satellite lead to many evaluations of
D/H (see e.g.  York and Rogerson 1976; Vidal--Madjar {\it et~al.}
1977; Laurent {\it et~al.} 1979; Ferlet {\it et~al.} 1980; York~1983;
Allen {\it et~al.}  1992) which also show dispersion around the above
York \& Rogerson (1973) value.  This dispersion has been recently
confirmed by IMAPS observations (Jenkins {\it et~al.} 1999; Sonneborn
{\it et~al.} 2000), indicating that the D/H ratio may vary by a factor
$\simeq 3$ in the solar neighborhood, i.e., within a few hundred
parsecs.

In this paper we present a new determination of the D/H ratio on the
line of sight to the nearby DA white dwarf (WD) G191-B2B based on
observations obtained with the Far Ultraviolet Spectroscopic Explorer
({\it FUSE}, Moos {\it et al.} 2000; Sahnow {\it et al.} 2000). This
paper is one of a series in this volume describing the first results
of the {\it FUSE} (D/H) program in the Local ISM (LISM). This program
and its results are summarized in the overview paper by Moos et
al. (2002).

Observing white dwarfs has many advantages over hot and cool stars for
studying the D/H ratio, as explained in Vidal-Madjar {\it et al.} (1998):
these targets can be chosen close to the Sun, in order to avoid a
complex line of sight structure, and in the high temperature range, so
that the interstellar absorption is superimposed on a smooth stellar
continuum. The risk of contamination by low column density H~{\sc i}
fluffs possibly present in the hot star winds (Gry, Lamers \&
Vidal--Madjar 1984) is negligible for WDs, and their hot continuum
offers the possibility of observing the numerous UV lines of N~{\sc i}
and especially O~{\sc i}, which is a very useful tracer of H~{\sc i} and
D~{\sc i}. The (D/H) ratio has already been measured toward four white
dwarfs, using the HST: G191-B2B (Lemoine {\it et~al.} 1996;
Vidal--Madjar {\it et~al.} 1998; Sahu {\it et~al.}  1999), HZ43A
(Landsman {\it et~al.} 1996), Sirius~B (H\'ebrard {\it et~al.} 1999)
and Feige~24 (Vennes {\it et~al.} 2000).  For HZ43A, Sirius~B and
Feige~24 the average D/H values obtained are compatible with the local
ISM (LISM) D/H determination (Linsky 1998) made within the Local
Interstellar Cloud (the LIC in which the sun is embedded), although in
the case of Sirius~B this compatibility is marginal.

In the case of G191-B2B, it was found that the line of sight comprises
one neutral region corresponding to the LIC, and two more ionized
absorbing components (Lemoine {\it et al.} 1996; Vidal-Madjar {\it et
al.} 1998). The average (D/H) ratio (defined as the ratio of the total
column densities of D~{\sc i} and H~{\sc i}) was found to be
(D/H)$=1.12\pm0.08\times10^{-5}$ (Vidal--Madjar {\it et~al.} 1998),
significantly different from the value measured toward Capella
(D/H)$_{\rm LIC}=1.5\pm0.1\times10^{-5}$ (Linsky 1998). The (D/H)
ratio measured toward G191-B2B has been contested by Sahu et
al. (1999), who used STIS data and concluded to the presence of two
interstellar components only, and to a D/H ratio compatible with that
observed toward Capella. The disagreement resides in the evaluation of
the total D~{\sc i} column density (Vidal-Madjar 2000; Sahu 2000), and
arises presumably because the \lya\ D line is saturated and the column
density is thus sensitive to the line profile. The number of
components assumed on the line of sight may also introduce differences
between the analyses of these groups (see Vidal-Madjar 2001 for a
detailed discussion).

In the present work we re-examine these issues, making use of high
quality {\it FUSE} and STIS observations of G191-B2B. We first measure
the \ndi , \noi\ and \nni\ column densities using the unsaturated lines
of these elements in the {\it FUSE} datasets, notably \lyb\ and \lyg\
for D~{\sc i} (Section~2). We then analyze the recent high quality STIS
observations in Section~3, and provide explicit evidence for the
presence of three absorbing components (at least) on the line of
sight. We also provide a refined estimate of the total H~{\sc i} column
density. All throughout this work, considerable effort has been put on
quantifying possible systematic uncertainties related to fixed-pattern
noise, detector artifacts, background uncertainties, wavelength
calibration and modeling of the stellar continua, as well as to the
velocity structure of the line of sight. In particular, we argue in
Section~3.3 that previous estimates of the total \nhi\ are subject to
a large systematic uncertainty related to the possible presence of
additional weak [\nhi$\leq10^{14}\,$cm$^{-2}$] hot ($T\sim10^5\,$K)
components. This effect may have a large impact on our understanding
of the observed variations of the (D/H) ratio in the ISM, as it may
affect other lines of sight, and is the subject of a companion paper
(Vidal-Madjar \& Ferlet 2002). We provide a summary of our results
and a short discussion in Section~4; an overall discussion of the {\it
FUSE} results is given by Moos {\it et al.} (2002).

\section{Column densities of D~{\sc i}, O~{\sc i} and N~{\sc i}
measured with {\it FUSE}}

\begin{figure}
\begin{center}
\psfig{file=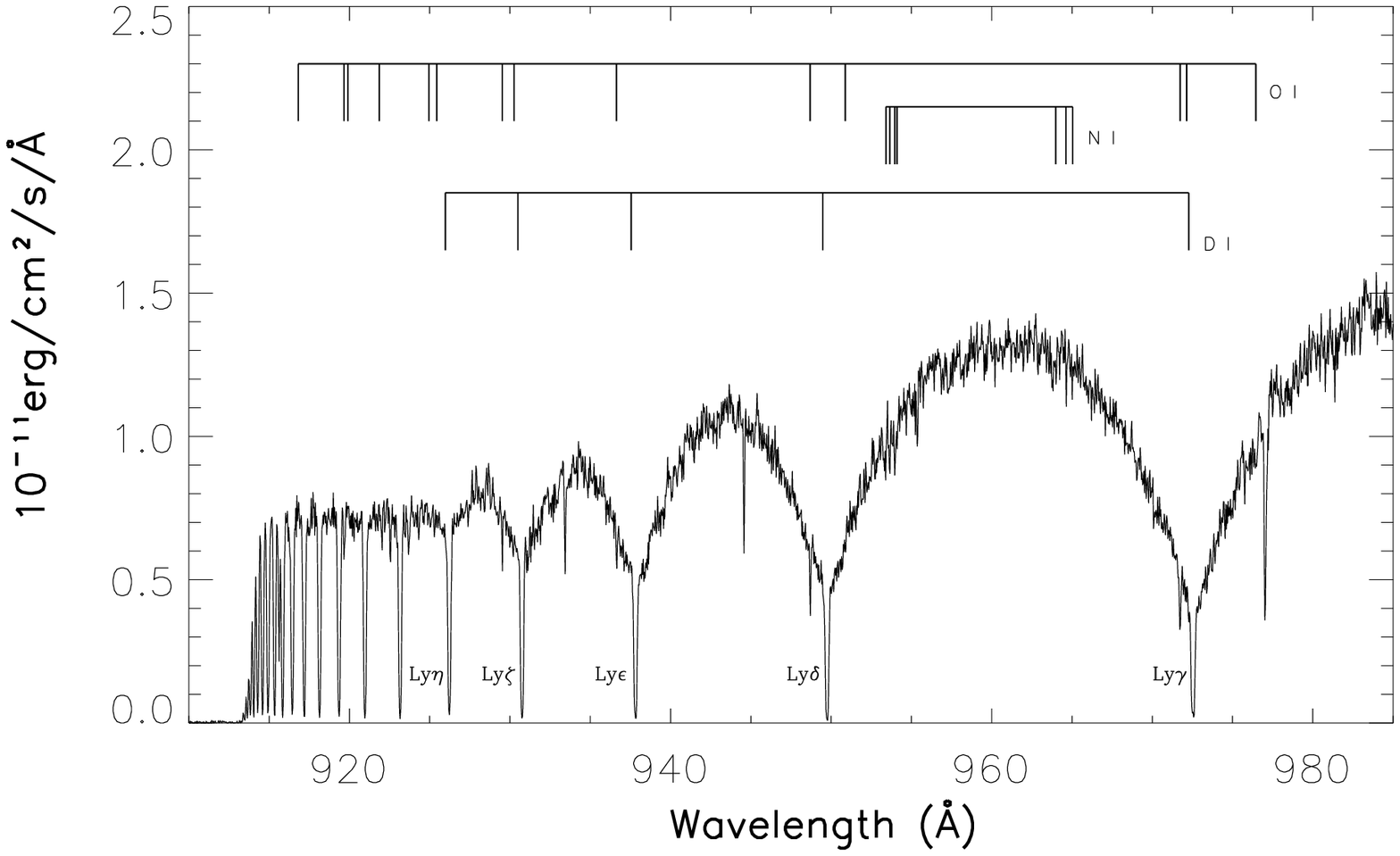,width=13cm}
\caption[]{{\it FUSE} HIRS spectrum of G191-B2B (segment SiC1B).  The
Lyman series is clearly seen down to the Lyman limit. The positions of
the D~{\sc i}, O~{\sc i} and N~{\sc i} lines considered in our study are
indicated.}
\label{FUSE_HIRS_1bsic}
\end{center}
\end{figure}

The {\it FUSE} instrument (Moos {\it et al.} 2000; Sahnow {\it et al.}
2000) gives access to the wavelength range 905--1187\AA\ on eight
detector segments, through three different entrance apertures (the
large, medium and narrow apertures -- LWRS, MDRS and HIRS,
respectively) and four channels (two LiF and two SiC). In particular,
it gives access to the higher order unsaturated Lyman lines of D~{\sc
i} and can thus offer a reliable estimate of \ndi. Similarly, many
weak lines of O~{\sc i} and N~{\sc i} are present in this bandpass, as
shown in Table~\ref{FUSE-lines}. The FUSE spectrum of G191-B2B in the
SiC1B channel observed with the HIRS aperture is shown in
Fig.~\ref{FUSE_HIRS_1bsic}, in which the lines of D~{\sc i}, N~{\sc i},
O~{\sc i} used in our study are indicated.

The white dwarf G191-B2B has been observed several times with {\it
FUSE} through the three different entrance apertures; the observation
log is given in Table~\ref{FUSE-GBB}. Since the target is point-like,
the resolutions obtained with these apertures should not be too
different, but there is a clear variation due to some off-axis effect
when moving from one aperture to another.  The main problem comes
from the geocoronal Lyman~$\beta$ emission whose strength increases
with the surface area of the aperture, and which blends with the
interstellar H~{\sc i} absorption. This leads us to employ different
approaches in the analysis of the data, depending upon the slit used,
as discussed below.

The observations were treated through the version 1.8.7 of the CALFUSE
pipeline. The data were then collected channel by channel by series of
subexposures. Due to the sensitivity of {\it FUSE} and the brightness
of the source, the data were obtained in the HISTOGRAM mode (a
spectral imaging mode, in which a two-dimensional spectrum was
accumulated onboard). Each subexposure has a S/N ratio that is high
enough to clearly see the strong interstellar or photospheric
features.  The different subexposures could then be easily aligned on
top of each other to compensate for the slow thermal drifts that
displace the wavelength scale from one subexposure to another. These
drifts never translates into more than $\pm5$ pixels.  A high S/N
spectrum is thus reconstructed in each channel and for each
aperture. This process preserves the ultimate {\it FUSE} resolution
and it also partly eliminates some of the detector's fixed pattern
noise by acting like a random FP-SPLIT procedure.

\begin{figure}
\begin{center}
\psfig{file=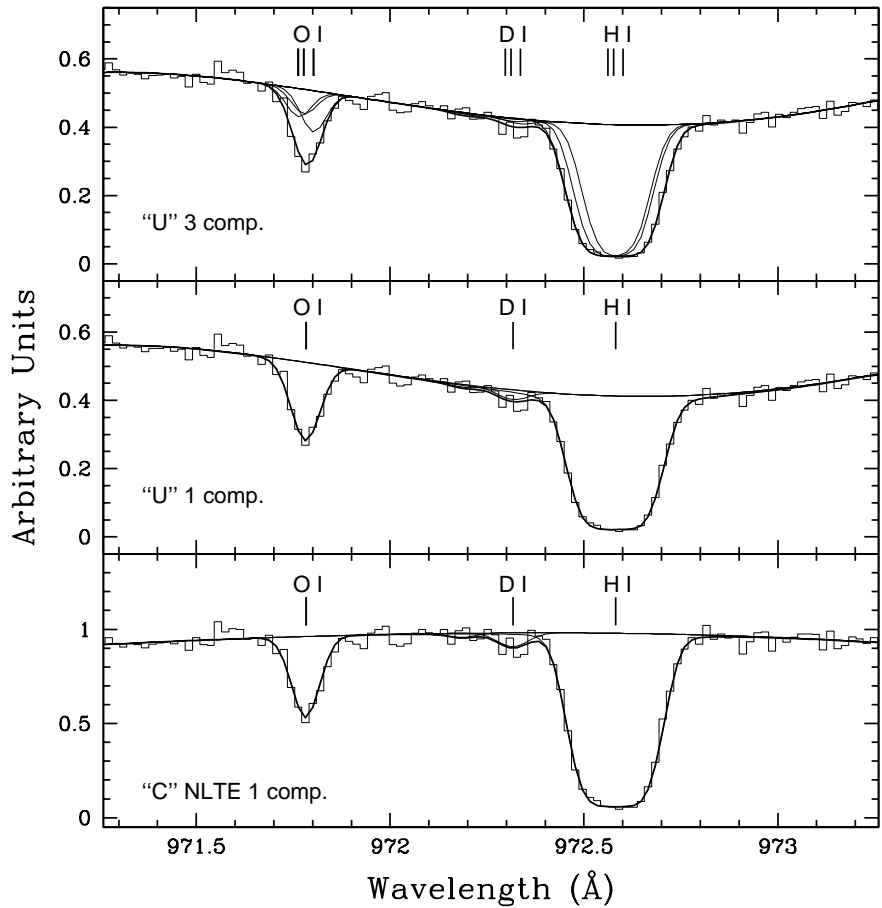,width=13cm}
\caption[]{{\it FUSE} MDRS spectrum of G191-B2B around \lyg\ (segment
SiC2A). The solid line in each panel shows the best fit obtained by a
simultaneous profile fitting of all H~{\sc i}, D~{\sc i}, O~{\sc i} and
N~{\sc i} lines of all {\it FUSE} datasets. Clearly these observations
do not resolve the multiple absorbers along the line of sight. In the
upper panel the continuum is interpolated by a smooth polynomial, and
three interstellar components corresponding to the velocity structure
of the line of sight derived from the higher resolution STIS
observations (see Section~3) were considered; in the middle panel, the
stellar continuum is also interpolated by a polynomial, while only one
absorbing component is considered. In the lower panel the stellar
continuum has been corrected by a theoretical NLTE stellar profile,
the residual continuum is modeled by a polynomial and only one
absorbing component has been assumed. See text for details.}
\label{FUSE_LyG}
\end{center}
\end{figure}

Since the {\it FUSE} sampling is $\sim10$~pixels per resolution
element, we rebinned the spectra by three. We measure column densities
by profile fitting of the various lines observed, using the code {\tt
Owens.f} developed by one of us (M.~L.) and the French {\it FUSE}
team. Our general procedure for data analysis is as follows. We split
each spectrum into a series of small sub-spectra centered on
absorption lines to be analyzed, whose width is a few \AA\ depending
on the line density and fit all lines contained in all sub-spectra
simultaneously. Between different sub-spectra, the zero-point velocity
offset is left to vary to compensate for the wavelength calibration
ripple effects in {\it FUSE} data; the Line Spread Function is
generally taken to be a Gaussian whose FWHM varies freely during the
fitting procedure and independently from sub-spectrum to
sub-spectrum. It is also checked that considering a double Gaussian
LSF with free widths, free amplitude ratio and zero pixel offset
between the two Gaussian components, does not change the results. The
background levels used for each line are those evaluated at the bottom
of the closest Lyman line (see Fig.~\ref{FUSE_LyG} for an example),
which leads to an upper limit for the derived column densities; fixing
the background level at the 0 value gives access to the lower limit of
the evaluated column densities. This uncertainty is taken into account
in our procedure; however {\it FUSE} is operating in first order
grating mounts and consequently the stray light level is generally
quite low, as can be seen in Fig.~\ref{FUSE_LyG} for \lyg.  This
analysis procedure allows to reduce systematic uncertainties tied to
fixed pattern noise effects or detector artifacts and uncertain
calibration, since different lines of the same elements observed
through different apertures and in different channels are fitted
simultaneously. For instance, if a given line is subject to an
artifact in the dataset, the \chid\ of the fit in the region around
this line, or equivalently the quality of the fit of this particular
line should stand out in the overall fit if the number of lines
analyzed is sufficiently large, as is the case here. Moreover, by
letting the unknown instrumental parameters vary freely, i.e., the LSF
shape and width, the background flux, the wavelength zero-point and
the shape of the continuum, and be optimized simultaneously with the
physical parameters that define the absorbers, one marginalizes the
final result over these unknowns. Therefore the errors attached to the
uncertainty on these above characteristics are included in the final
errors quoted for the column densities. Finally, all lines are also
analyzed one by one, {\it i.e.} each independently of all others, in
order to check for overall consistency. A detailed discussion of this
procedure and survey of possible systematic errors tied to {\it FUSE}
data is given in a companion paper by H\'ebrard {\it et al.} (2002).

\subsection{O~{\sc i} and N~{\sc i}}

Since {\it FUSE} cannot resolve the velocity structure of the line of
sight (see Section~3), we have used only unsaturated lines of D~{\sc
i}, N~{\sc i} and O~{\sc i} and assumed only one interstellar
component to be present. Therefore, all measured column densities in
this Section are integrated over the line of sight. It has been
verified that using three interstellar components in the fit of the
{\it FUSE} did not change the conclusions as long as only
non-saturated lines were considered; an example of single {\it vs}
multiple component fits is shown in Fig.~\ref{FUSE_LyG}. 

Note that HST observations cannot give a reliable measurement of the
O~{\sc i} column density without a complete knowledge of the velocity
and broadening structure of the line of sight since the only
absorption line available at 1302\AA\ is strongly saturated. In
contrast the N~{\sc i} triplet at 1200\AA\ in the GHRS or STIS
bandpass is only moderately saturated and may provide a precise
estimate of the total \nni. The various measurements of \nni\ and
\noi\ are listed in Table~\ref{OI_pub}.

\begin{figure}
\begin{center}
\psfig{file=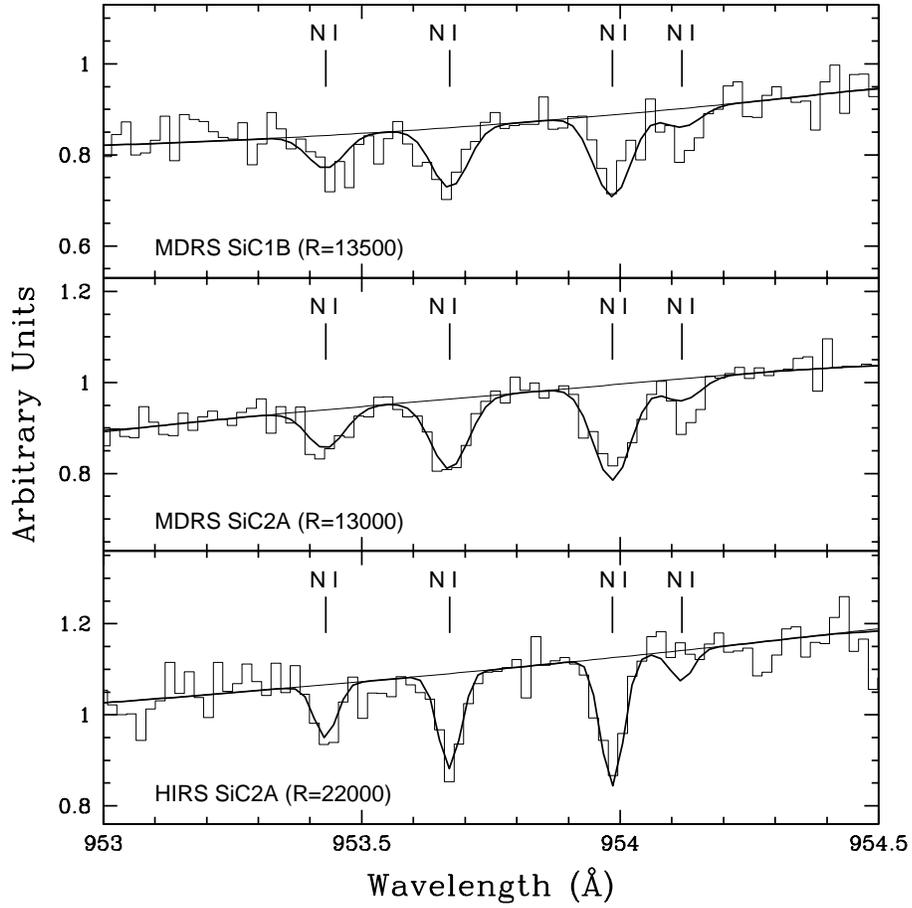,width=13cm}
\caption[]{ {\it FUSE} spectra of G191-B2B around the N~{\sc i}
multiplet at 954\AA. The aperture, segment and resolving power $R$
corresponding to each dataset are indicated on the figure. The line at
954.1\AA\ was excluded from the profile fitting since it is likely
contaminated by a photospheric feature, as indicated by theoretical
modeling of the stellar continuum, and as shown clearly in the upper
and middle panels; in the lower panel, the apparent absence of this
line may result from a statistical fluctuation or from a detector
feature. The higher resolution of the HIRS aperture data is apparent.}
\label{FUSE_NI}
\end{center}
\end{figure}

 Since the contamination of interstellar lines by weak photospheric
lines is a likely possibility in the {\it FUSE} spectral range due to
the high density of atomic lines in this bandpass, we excluded from
the fit all lines that presented an obviously excessive equivalent
width or column density relative to the others. As an example as shown
on Figure~\ref{FUSE_NI} the N~{\sc i} line at 954.1042\AA\ is very
probably blended with a photospheric line which shows up at nearly the
same wavelength in the NLTE spectrum calculated with the
TLUSTY/SYNSPEC codes (described in Section~3). Such blending becomes
particularly important in the case of G191-B2B because the total O~{\sc
i} or N~{\sc i} column densities are small, and their absorptions are
relatively weak features which may be more easily blended with
undetected photospheric ones.  This is much less important for the
other {\it FUSE} (D/H) targets (see the synthesis of {\it FUSE}
results by Moos {\it et al.}~2002, and the individual line of sights
studies by Friedman {\it et al.}~2002, H\'ebrard {\it et al.}~2002,
Lehner {\it et al.}~2002, Sonneborn {\it et al.}~2002 and Wood {\it et
al.}~2002) which present larger column densities, except for the white
dwarf HZ43A (Kruk {\it et al.}~2002) which presents a pure H
atmosphere (Barstow, Holberg \& Koester 1995; Dupuis {\it et al.}
1998) and thus no possible blend with photospheric lines.  However
note that this line selection process may induce in the case of
G191-B2B a slight underestimation of the total O~{\sc i} or N~{\sc i}
column densities.

The O~{\sc i} and N~{\sc i} lines used in this study are listed
in Table~\ref{FUSE-lines}. Some of them were rejected either because
they are saturated (marked ``strong'' in Table~\ref{FUSE-lines}), or
because of a blend with a photospheric feature (marked ``blend''), or
because they are too weak and subject to systematic effects of the
noise (marked ``weak'').  Some were kept in one channel but not in an
other if a nearby detector defect was identified. For this reason,
having the same spectrum observed in different independent channels
was extremely helpful.

By following the general method depicted above, the final column
density determinations are~:

\begin{center}
log~$N$(O~{\sc i})$_{\rm tot}$~=~14.86~($\pm0.05$)~($\pm{0.04}$) \\ 
log~$N$(N~{\sc i})$_{\rm tot}$~=~13.87~($\pm0.05$)~($\pm{0.05}$)  
\end{center}
where the errors are first statistical (2$\sigma$) and second
systematic (\cl ). The statistical errors were evaluated through the
$\Delta\chi^2$ technique, an example of which is given for the \ndi\
determination to follow.  However, one important comment is of order
at this point. The total \chid\ for all spectral lines fitted
simultaneously (H~{\sc i}, D~{\sc i}, O~{\sc i} and N~{\sc i}) in the case
of the HIRS observations is equal to 3211.44 for 1843 degrees of
freedom (d.o.f.), even though the fit appears very satisfactory by
eye. This \chid\ value lies well above the upper 95.5\% confidence
level limit for statistical fluctuation at 1966.71. We interpret this
large \chid\ value as a result of a systematic underestimate of the
magnitude of the individual pixel errors by the instrument. In order
to quantify this uncertainty, we have completed low order polynomial
fits over relatively smooth and flat continuum sections close to each
of the spectral lines analyzed. These simple fits gave a $\chi^2$ of
688 for 500 degrees of freedom, again above the 95.5\% c.l. upper
limit at 563.9. This indicates that indeed the individual pixel errors
are underestimated by instrument, probably due to the presence of
fixed pattern noise in the data. The discrepancy in terms of ratio of
measured to expected \chid\ is not as strong as for the best-fit
\chid\ values, but this may be due to differences in the error levels
for pixels located on the continuum and those located at the bottom of
the saturated Lyman lines.

In order to compensate for these effects, we have thus decided to
rescale all \chid\ measured, which amounts to an overall rescaling of
the individual pixel error estimates. In particular, we have chosen to
divide all \chid\ by $1.84=3211.44/1741.64$, where $1741.64$
corresponds to the lowest expected \chid\ at the 95.5\% confidence
level for 1843 d.o.f. This rescaling rescales (increases) the error
bars on the physical parameters that we derive since the
$\Delta\chi^2$ contour levels are also rescaled. In effect, in order
to derive the $3\sigma$ error bar around a single parameter minimum
value, we search the values of this parameter that lead to an increase
in (non-rescaled) $\chi^2$ of $9\times1.84=16.6$ instead of $9$. Our
choice of rescaling is thus very conservative with respect to the
final error bars measured.  All presented FUSE statistical errors will
be evaluated following this method.

  The systematic errors quoted above for O~{\sc i} and N~{\sc i}
reflect the range of values obtained for the best-fit solutions for
different apertures, given in Table~\ref{OI_pub}. Note that in this
table, the errors quoted include the satistical error plus part of the
above systematic errors, since it accounts for the differences
measured between different observations, different channels and
different spectral ranges where lines of the species concerned are
seen, but observed through the same aperture (see the discussion in
H\'ebrard {\it et al.}~2002).  Note that the total O~{\sc i} or
N~{\sc i} column density evaluations made with the LWRS+MDRS are in
both cases lower than those made with the other apertures.  We do not
know the cause of this apparent systematic effect, but included it in
the overall uncertainty on our result.

The quadratic sum of statistical and systematic uncertainties then
leads to:

\begin{center}
   log $N$(O~{\sc i})$_{\rm tot}$ = 14.86 ($\pm0.07$) \\
   log $N$(N~{\sc i})$_{\rm tot}$ = 13.87 ($\pm0.07$) 
\end{center}

\subsection{Measurement of \ndi}

In the {\it FUSE} bandpass, D~{\sc i} is clearly detected at Lyman
$\beta$, weakly at Lyman $\gamma$, and not at Lyman $\delta$. We list
in Table~\ref{DI_pub} the different estimates of the total column
density of D~{\sc i} that have been obtained from HST observations of
\lya , and which we measure using the present {\it FUSE} data. Here as
well, we assumed that only one component is present on the line of
sight since the D~{\sc i} lines are optically thin. Again we checked
the impact of a more complex line of sight structure and found it to
be negligible, as expected. We carried out the same investigations of
systematic effects as for O~{\sc i} and N~{\sc i}, using the approach
detailed above and in H\'ebrard {\it et al.}  (2002).

\begin{figure}
\center
\psfig{file=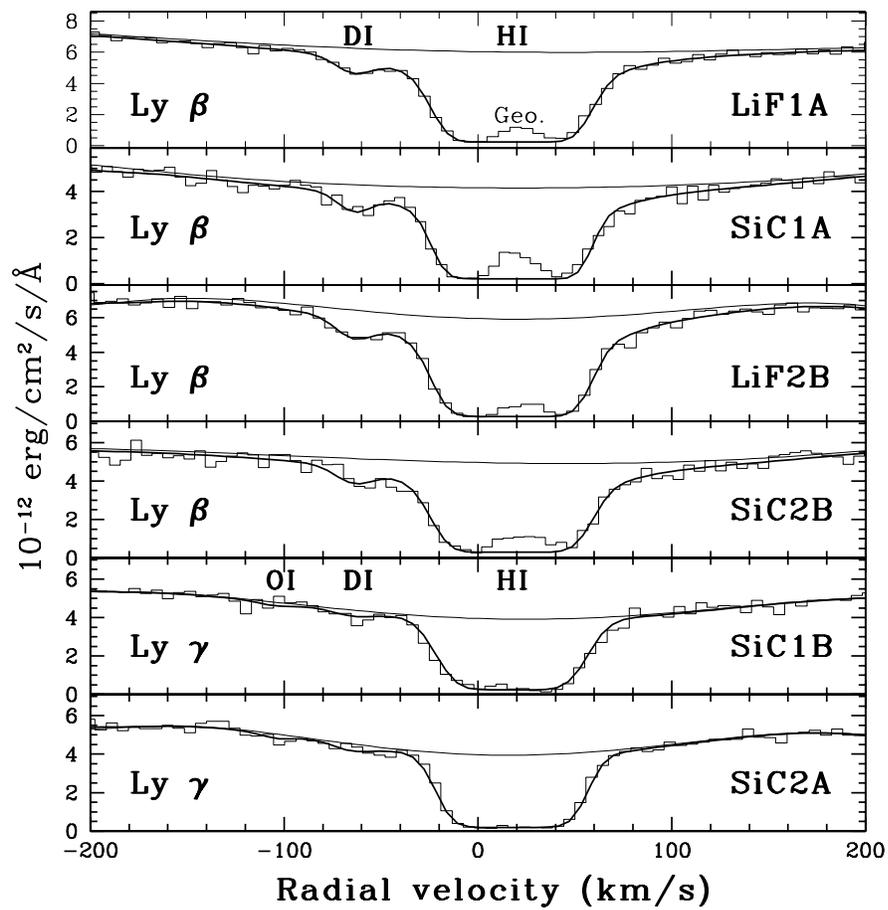,width=13cm}
\caption{{\it FUSE} HIRS spectra of \lyb\ and \lyg, recorded on the
segments indicated in the figure. The solid line shows the final
best-fit solution to all lines of H~{\sc i}, D~{\sc i}, N~{\sc i} and
O~{\sc i} in all HIRS data fitted simultaneously. The weak emission
line at the bottom of the Lyman lines is due to H~{\sc i} geocoronal
emission; the pixels affected by this emission are not considered in
the fit. The stellar continuum is interpolated by a smooth
polynomial. See text for details.}
\label{AAS_fig2}
\end{figure}

There is however one essential difference between D~{\sc i} and O~{\sc
i} or N~{\sc i} in terms of systematics. Namely the continuum to the
D~{\sc i} absorption is provided by the blue wing of the corresponding
H~{\sc i} absorption, while the continuum around the O~{\sc i} and N~{\sc
i} is very smooth. It is thus necessary to measure the possible
systematic tied to the estimate of the stellar continuum on the
measurement of \ndi. In order to do so, we have measured \ndi\ with
and without beforehand correction of the data by a theoretical NLTE
stellar continuum (described in Section~3). Some corresponding fits
are shown in Fig.~\ref{FUSE_LyG}, and the best-fit to the HIRS data is
shown in Fig.~\ref{AAS_fig2}. The effect of any assumption on the
stellar continuum was found to be negligible as compared to the
statistical uncertainty.

  However, we have found that the value of \ndi\ measured if the
H~{\sc i} lines are excluded from the fit and their blue wing modeled
by a polynomial is higher than when the H~{\sc i} are included in the
fit. This systematic effect may result from the following: in the
latter (standard) approach, where both D~{\sc i} and H~{\sc i} are
fitted together, the position of the D~{\sc i} absorption is tied to
that of H~{\sc i}, i.e., they have the same radial velocity and their
zero point wavelength offsets are the same as both lines appear in the
same sub-spectrum. However, in the former approach, the velocity of
D~{\sc i} is not subject to this constraint. These two approaches are
denoted in Table~\ref{DI_pub} as ``D~{\sc i} \& H~{\sc i}'' (standard
approach) and ``D~{\sc i}, no H~{\sc i}'' for the other one. One
should note that a discrepancy in radial velocity between the D~{\sc
i} line and its counterpart H~{\sc i} could be attributed to small
scale ripples in the {\it FUSE} wavelength calibration or to the
presence of weak H~{\sc i} absorbers which would shift the position of
the H~{\sc i} line with respect to that of the D~{\sc i} line in which
these absorbers would not be felt. In this respect the approach in
which D~{\sc i} is fitted independently of H~{\sc i} seems more
adequate; however the interpolation of the blue wing of the H~{\sc i}
line by a polynomial is not always straightforward.  We should mention
that this systematic effect is not fully understood and does not seem
to be always present on different lines of sight. In any case it is
included in our estimate of the systematic error.

\begin{figure}
\center
\psfig{file=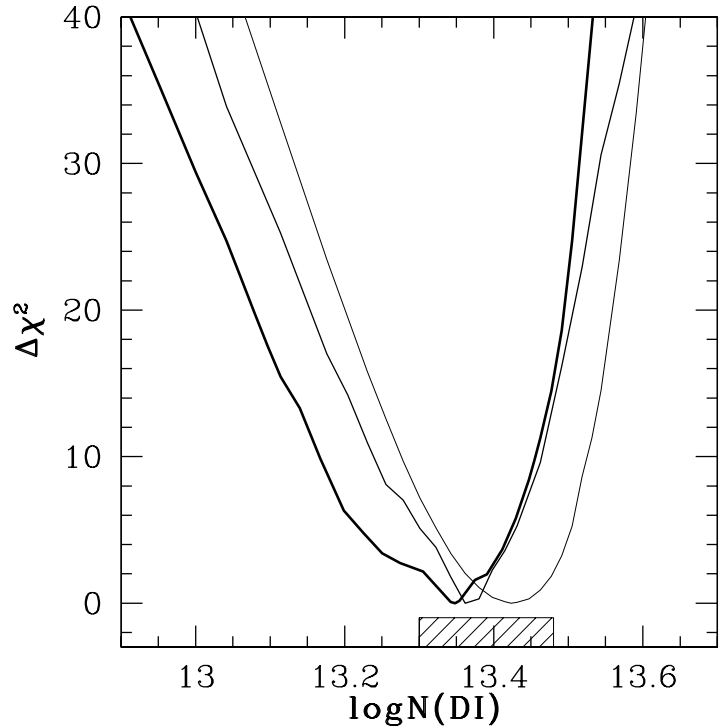,height=10cm}
\caption{Curve of $\Delta\chi^2$ deviations around the best fit \chid\
as a function of $\log$\ndi. Only the curve for the HIRS data is shown
here; statistical error bars for other aperture data were obtained
using the same method. The $\Delta$\chid\ values shown here have been
rescaled to compensate for uncertainties in the individual pixel
errors (see text). The various curves correspond to different models
for the fit: in thick line, the D~{\sc i} and H~{\sc i} are fitted
simultaneously with a LSF modeled as a single gaussian with free FWHM;
in intermediate thickness, the same approach but with a double
gaussian LSF, with free amplitude ratio and FWHMs; in thin line, H~{\sc
i} is excluded from the fit, the continuum to the D~{\sc i} absorption
is modeled by a polynomial, and the LSF is a simple gaussian. The
curvature of these curves give the the statistical errors while their
relative shifts give an estimate of the overall systematic
uncertainty related to the different models. The hatched area shows
the final 95.5\% confidence level error for $\log$\ndi\ measured using
the HIRS data.}
\label{AAS_fig3}
\end{figure}

Finally a small additional systematic effect is found to be attached
to the uncertainty in the shape of the LSF, i.e. whether it is a
single gaussian or a double gaussian with wide wings (indicated as
``double LSF'' in Table~\ref{DI_pub}). As mentioned previously, a
single gaussian LSF has a free FWHM in the fit, while a double
gaussian LSF has free amplitude ratio between both components, free
FWHM for each, but separation between both components fixed to zero
pixel. The impact of the shape of the LSF on the measured \ndi\ is
shown in Figure~\ref{AAS_fig3} where the $\Delta \chi ^2$ is plotted
as a function of $\log$\ndi. In practice this curve is calculated by
fixing $\log$\ndi\ to a given value, finding the best fit \chid\ for
this value with all other parameters free, and plotting the difference
between this \chid\ and the best \chid\ obtained for all possible
values of $\log$\ndi. In this figure, the curvature of a
$\Delta\chi^2$ curve gives the statistical error, as usual, while the
relative shift between two curves corresponding to two different sets
of assumptions is interpreted as an estimate of the $1\sigma$
systematic error tied to the uncertainty on the assumptions. 

We thus conclude that the total D~{\sc i} column density on this
line of sight measured using the HIRS data is:

\begin{center}
$\log N({\rm D}$~{\sc i}$)_{\rm tot}^{\rm HIRS} = 13.39 \pm{0.07}
\pm0.06 $\\
\end{center}
where the errors are first statistical (2$\sigma$) and second
systematics (\cl).

Combining these errors leads to:
\begin{center}
$\log N({\rm D}$~{\sc i}$)_{\rm tot}^{\rm HIRS} = 13.39 \pm{0.09}$
\end{center}

As can be seen from the comparison between the evaluations of \ndi\
made for the different apertures, there is no significant systematic
effect tied to the aperture, contrary to the analysis for O~{\sc i}
and N~{\sc i}. The average \ndi\ measured through the different
apertures then reads:
\begin{center}
$\log N({\rm D}$~{\sc i}$)_{\rm tot} = 13.40 \pm0.07$
\end{center}

This value agrees with all previously measured values of \ndi\, shown
in Table~\ref{DI_pub}, except with that of Sahu {\it et al.} (1999),
using the STIS-Ech\#1 data, which gave
$\log$\ndi$=13.55^{+0.07}_{-0.08}$. It thus appears that the
discrepancy between this value and the others should be attributed to
the dataset used, in agreement with the conclusion of Vidal-Madjar
(2001), and the debate around the value of \ndi\ is now settled.

\section{The velocity structure of the line of sight and \nhi}

In this Section we analyze new high resolution high signal-to-noise
STIS data of G191-B2B in order to determine the number of components
on the line of sight and provide a new estimate of the total neutral
hydrogen column density. As mentioned in Section~1, there exists a
controversy in the literature about the total number of absorbing
components, as Sahu {\it et al.} (1999) claimed to see two only,
whereas Vidal-Madjar {\it et al.} (1998) claimed that three at least
were present. In this Section, we present concrete evidence in favor
of the latter. Previous measurements of \nhi\ are compiled in
Table~\ref{HI_pub}; these values scatter around a (non-weighted) mean
$\log$\nhi$=18.34\pm0.03$ (the dispersion corresponds to the
non-weighted dispersion of the individual measurements around the
mean). Individual measurements tend not to agree with each other, and
therefore a new independent estimate is useful. Moreover we argue in
this Section that the total \nhi\ toward G191-B2B is, as a matter of
fact, much less well known than previously thought, since additional
weak hot absorbing components may strongly affect the column density
estimate from Lyman $\alpha$. The new STIS data analyzed here
represent about three times the total exposure time cumulated hitherto
toward G191-B2B.

All STIS observations were extracted following the method of Howk and
Sembach (2000). We selected spectral lines corresponding to N~{\sc i},
O~{\sc i}, Si~{\sc ii}, Si~{\sc iii}, C~{\sc ii}, S~{\sc ii}, S~{\sc iii} in
the E140H echelle configuration and Fe~{\sc ii} lines in the E230H one,
both at $R\simeq90,000$ resolving power, as reported in
Table~\ref{STISlines}.  All these species are seen in one or more
spectral lines, strong and/or weak, and cover a wide range of atomic
masses, so that this data should provide strong constraints on the
structure of the line of sight as well as on the component to
component physical state and thermal and non-thermal broadening of
each.

In order to verify that no photospheric line is blended with one of
the interstellar lines analyzed, we have computed a synthetic spectrum
of G191-B2B which includes all species observed in the atmosphere of
G191-B2B, i.e., C, N, O, Si, P, S, Fe, and Ni (Lanz {\it et al.} 1996;
Vennes {\it et al.} 1996).  We employ the program TLUSTY developed by
Hubeny \& Lanz (1995) to compute a NLTE metal line-blanketed model by
using the atmospheric parameters of Barstow {\it et al.} (1998), i.e.,
$\log g= 7.4$ and $T_{\rm eff}=54\,000$K, and the abundances
determined by Barstow {\it et al.} (2001).  From the comparison of the
observations with the model calculation it was possible to select the
lines where no or very weak photospheric features are coincident with
the interstellar ones, as for the {\it FUSE} data. Furthermore it is
possible that some photospheric lines are present but not predicted by
the model; we thus fitted simultaneously as many different spectral
lines of each species as available, in order to identify and minimize
the effect of unpredicted photospheric lines, as was done for the
analysis of the {\it FUSE} data. This approach also reduces the
possible impact of instrumental defects which may be present in some
area of the detector but not in others.

  Following the general method of analysis described in the previous
Section, we split the STIS data into sub-spectra, each of them
centered on one spectral line and having typical width
$\sim0.3$\AA. All sub-spectra are then fitted simultaneously. The
wavelength zero point offsets are left free to vary during the
fit. Note that shifts in the wavelength calibration from region to
region could still be present even after a careful wavelength scale
calibration. As an example, we detected an erroneous 4\kms wavelength
shift between the GHRS Si~{\sc iii} line position and that in the STIS
data. After comparison with GHRS first order grating observations, we
confirmed that the STIS calibration was correct, and the GHRS
calibration erroneous. We found that the average relative shift
between two spectral regions in the STIS data is
$\simeq-0.1\pm1.1$\kms\ ($2\sigma$ error). An {\it rms} error of
0.5\kms\ in the STIS calibration is indeed compatible with the
$\simeq$3.3\kms\ resolution of the instrument.  The stellar continua
in each region are interpolated by low order polynomials (except for
\lya\ where the theoretical continuum is used in some cases, see
below), and the background level in each region is determined using
the closest saturated line. The typical background flux is
$-2.6\pm2.4\%$ (2$\sigma$ error) of the continuum for the E140H data,
and $-0.10\pm4.2\%$ (2$\sigma$ error) for the E230H data. Finally in
some cases (see below for details), the LSF is left free to vary and
corresponds either to a simple gaussian or to a double gaussian.
Again, we emphasize that the uncertainty on these instrumental
characteristics is contained in the final error bars given below,
which we determine using a $\Delta\chi^2$ method, since the
corresponding parameters are left free to vary during the fit.

In the case of the STIS observations made with the E140H and E230H
echelle modes, we used the tabulated LSF, corresponding to the slit
used for these observations (0.2$\times$0.2 arc sec).  This is an
important issue because these LSFs possess wings that could have some
impact on the precise determination of the physical parameters we are
aiming at. To test the quality of the fits we also tried a single
gaussian LSF with free FWHM.  We found excellent fits with this freely
variable width single gaussian LSF for both echelle datasets. We find
for the E140H an average FWHM$=2.91(\pm 0.44)$~pixel (2$\sigma$) and
for E230H, FWHM$=2.38(\pm 0.47)$~pixel (2$\sigma$). These
determinations are certainly compatible with the tabulated LSF in
terms of ``average'' widths, but are different in terms of shape. They
correspond to 80,000 and 100,000 resolving power respectively for the
E140H and E230H spectra.  Finally, we also tested double gaussian LSFs
but found no significant improvement.  We thus decided to present
results with both types of LSFs (free gaussians and tabulated) in
order to show the stability and robustness of our column density
determinations. As we will see, such a change has negligible impact on
the derived total H~{\sc i} column density.

Following the analysis of the {\it FUSE} data, we evaluate the
accuracy of the intrument estimate of the individual pixel errors by
completing low order polynomial fits over relatively smooth and flat
continuum sections close to each of the fitted spectral lines. These
simple fits give a total \chid =921.37 (sum of all \chid\ in the
different portions of continuum analyzed) for 661 degrees of freedom,
which lies well above the upper 95.5\% confidence level limit of 723.9
for a $\chi^2$ distribution with 661 d.o.f. Therefore we rescale all
\chid\ by a common factor $921.37/600.63 = 1.53$, where 600.63
corresponds to the lower 95.5\% confidence level limit for 661 d.o.f.

Finally, in the course of the analysis, the physical characteristics
of each absorbing component (velocity $v$, temperature $T$ and
micro-turbulent broadening $\xi$) are determined; however we will not
provide a detailed analysis of these characteristics and their
uncertainties since we are primarily interested in the total number of
absorbing components and the total neutral hydrogen column density.

\subsection{Number of absorbing components}

In order to settle the debate on the total number of absorbing
components, we have studied two and three component solutions. Simple
eye inspection of the data indicates that at least two components are
present on the line of sight; as mentioned previously, the red
component can be identified with the LIC, a moderately neutral region,
while the blue component is clearly more ionized, and, as we argue,
more complex than a single absorber (see Vidal-Madjar {\it et al.}
1998 and Sahu {\it et al.} 1999).

\begin{figure}
\begin{center}
\psfig{file=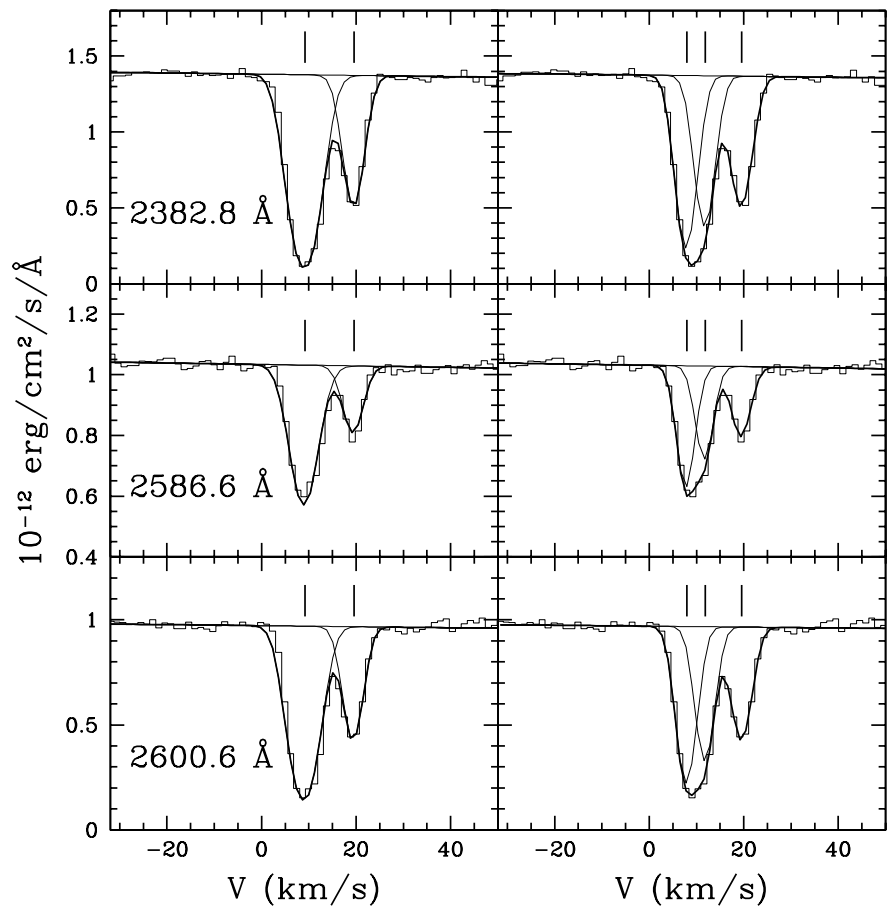,width=13cm}
\caption[]{ The STIS data covering the three Fe~{\sc ii} lines observed
with the E230H Echelle grating. All fits shown here were made using a
freely varying single gaussian LSF; the Fe~{\sc ii} lines were fitted
simultaneously with all other species including Lyman~$\alpha$ (see
text).  The left panels show the fits with two absorbing components
only, and the right panels show the fits using three components. One
can clearly see the same asymmetry of the bluer component {\it B} in
all Fe~{\sc ii} lines which reveals the presence of complex
substructure in this component, hence the need for 3 absorbers in
total. The $\chi^2$ corresponding to these fits is $227.9/139$ for the
two component solution, and $123.3/136$ for the three component
solution; the gain is clear and comforts the visual impression. The
error bars on each pixel are of order of 2\% of the continuum level.}
\label{Fe_2comp_3comp}
\end{center}
\end{figure}

  The result of this analysis is summarized in Table~\ref{chi2}, which
gives the \chid\ values obtained for the fits of various spectral
regions, and for all regions (excluding \lya), for two and three
component solutions, using either the tabulated STIS LSF or a freely
varying single gaussian LSF.  It is clear that the improvement in
$\chi^2$ when going from two to three components is extremely high
whichever LSF is used.  This is particularly true for Fe~{\sc ii} in
which case the improvement can be seen directly on
Figure~\ref{Fe_2comp_3comp}.  Indeed if one looks closely at the
Fe~{\sc ii} lines, one can see an asymmetry of the line profile of the
bluer component which is common to all Fe~{\sc ii} lines. This
constitutes clear evidence for the presence of three absorbing
components and not two. One sees this third component in Fe~{\sc ii}
and not in other lines since Fe~{\sc ii} is the heaviest species, and
its line widths are thus the smallest.  Moreover its lines have been
observed through the echelle presenting the higher resolving power
(100,000). This suffices to reveal the double structure of component
B: components B1 and B2 are separated by $3.8\pm0.5$
\kms\ (2$\sigma$), {\it i.e.} of the order of the E140H spectral
resolution but more than that of the E230H. The radial velocities of
the three components is estimated as $v_{\rm B1}=7.7\pm0.5$\kms,
$v_{\rm B2}=11.5\pm0.5$\kms and $v_{\rm LIC}=19.4\pm0.5$\kms, with
$1\sigma$ errors due to the wavelength calibration. The LIC velocity
is in perfect agreement with its projected velocity on the line of
sight.

\begin{figure}
\begin{center}
\psfig{file=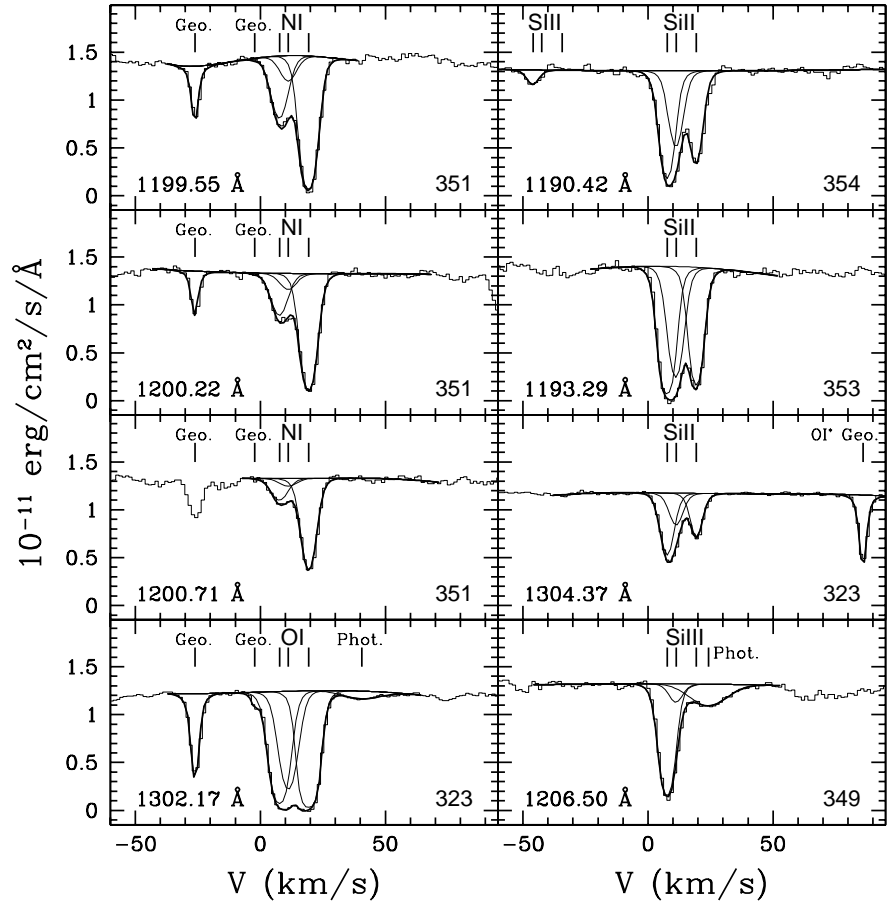,width=13cm}
\caption[]{The STIS data with the best-fit three component solution
using the tabulated STIS LSF; only 8 sub-spectra out of 19 in total
are shown here, and are labeled with the central spectral line
wavelength and Echelle order: the N~{\sc i} 1200\AA\ triplet along with
the nearby geocoronal absorptions (marked ``Geo.'', in two different
locations because two data sets taken at two different epochs are here
averaged), the O~{\sc i} line with the corresponding geocoronal O~{\sc
i} absorptions and a nearby photospheric feature (noted ``Phot.''),
three of the Si~{\sc ii} lines with either the nearby S~{\sc iii}
feature or the O~{\sc i}$^*$ geocoronal absorption and the Si~{\sc iii}
line with an additional broad Si~{\sc iii} photospheric line located at
the same velocity shift as the N~{\sc v} photospheric line (which was
fitted simultaneously). Note also that the S~{\sc iii} lines clearly
detected in two different orders near 1190.2\AA\ are also spectrally
well located. The fact that they show up only in the bluest component
B1 is an additional reason why 3 and not 2 components are needed
along that line of sight.}
\label{STIS_3comp_Fig3}
\end{center}
\end{figure}

One can also see further evidence for the double structure B1-B2 in
the S~{\sc iii} 1190.2\AA\ line shown in the upper right panel of
Fig.~\ref{STIS_3comp_Fig3}. This line is seen only in the bluest
component B1, which reflects the difference in ionization between the
various components; its spectral position is quite precisely known as
it is located close to the Si~{\sc ii} line at 1190.4\AA\ and is seen in
two different spectral orders (see Table~\ref{STISlines}). Note that
the position of most lines is actually very well controled, which is
important to determine the number of absorbing components. In effect,
the sharp geocoronal lines of N~{\sc i}, O~{\sc i} and O~{\sc i}$^*$ are
in several locations (see Fig.~\ref{STIS_3comp_Fig3}), notably close
to interstellar lines of N~{\sc i}, O~{\sc i} and Si~{\sc ii}. Other
Si~{\sc ii} lines or Fe~{\sc ii} lines are well constrained since in
each of these species the absorption due to the LIC is sharp and
pronounced, and serves as a useful spectral reference.

In order to test the hypothesis that the presence of a third component
is not required by the fit, we perform an $F-$test.  This test uses
the Fisher-Snedecor law which describes the probability distribution
of the $\chi^2$ ratio. Here we test the probability that the decrease
of the $\chi^2$ due to the inclusion of a third component could be due
to the increase of free parameters and not to actual information
contained in the data.  The result is shown species by species and for
all species fitted simultaneously in Table~\ref{chi2}, and gives a
probability $\leq10^{-4}$ that this third component is not required by
the data. The $F-$tests performed species by species also confirm our
previous impression that Si~{\sc ii} and Fe~{\sc ii} are the two most
sensitive species to the presence of this component. We also note that
the need for a third component is present whichever LSF is used.

We note that the \chid\ values corresponding to the use of the
tabulated STIS LSF are too high, as they all lie above the upper
95.5\% confidence level limit for statistical fluctuation (an except
being Fe~{\sc ii} for three components). In contrast, the \chid\ values
for a freely varying single gaussian LSF are satisfactory, since only
O~{\sc i} and Si~{\sc ii} stand out with \chid\ beyond the 95.5\%
confidence level for the three component solution. Their high \chid\
propagates into the total \chid\ summed over all windows. However the
Si~{\sc ii} 1190\AA, 1193\AA\ and O~{\sc i} 1302\AA\ lines are
saturated, and small errors in the estimates of the error bars at the
bottom of the line could potentially explain large differences in
\chid. In effect our procedure of rescaling the $\chi^2$ amounts to
rescaling all error bars by a common factor, which may not be correct
for pixels at the bottom of saturated lines for the following
reason. Error bars on flux values arise as a combination of background
noise, fixed pattern noise and Poisson noise. Background noise
dominates for pixels with flux values close to zero, while for high
S/N data Poisson noise or fixed pattern noise would dominate in the
continuum.  If the errors in the estimation of the noise array for the
background component and for the continuum component (fixed pattern or
Poisson) are different, then the procedure of rescaling error bars
with a common factor is, strictly speaking, incorrect. Finally, since
this rescaling factor was estimated from the continuum, this rescaling
should be correct for all pixels whose error is not dominated by the
other noise component (background). We also note that this O~{\sc i}
line has a complex profile, with two geocoronal and one nearby
photospheric feature (see Figure~\ref{STIS_3comp_Fig3}). Nevertheless,
we decided to keep this line in our analysis as it adds important
constraints on the B1 and LIC component T and $\xi$
parameter evaluations.

Using the $F-$test, we can determine the probability that the use of
the freely varying Gaussian LSF is not required (instead of using the
STIS LSF). In all cases, it is found that this probability is smaller
than $10^{-4}$, and therefore we conclude that the gaussian LSF with a
width that varies from spectral region to spectral region is required
to fit the STIS data.  This result may be unexpected but it can find a
simple explanation in the fact that the tabulated STIS LSF is an
average LSF given over the whole spectral range, while our gaussian
LSF has a FWHM that varies with wavelength.  For standard datasets it
is probably sufficient to use the STIS LSF but the very high quality
of the present data necessitates higher order corrections to this
LSF. In particular we noted a slight broadening of the LSF near the
order edges.  Furthermore, the wings of the tabulated LSF are not so
important and the single gaussian LSFs are sufficient to properly fit
the data. 

We thus conclude that three absorbers are present on the line of sight
to G191-B2B.

\subsection{\lya}

The background level is directly measured at the bottom of the
strongly saturated Lyman~$\alpha$ line. It is found to be of the order
of $-1.0(\pm1.0)\%$ (2$\sigma$) of the nearby continuum, i.e. within a
few percent.  Variations in the photospheric continuum over the
interstellar Lyman~$\alpha$ line presents an additional difficulty.
Indeed the instrument sensitivity over this spectral region is
evaluated by using model spectra of white dwarfs and comparing them to
the data, and one of the calibration WDs used is G191-B2B itself.
Therefore it is difficult to separate the instrumental effects from
the photospheric intrinsic profile in the spectrum processed through
the STIS pipeline.  For this reason, we follow the approach detailed
in Vidal--Madjar {\it et al.} (1998) and further discussed by
Vidal--Madjar (2000), and evaluate simultaneously all parameters and
the stellar continuum, which we model in two different ways. In the
first approach, we model this continuum as a polynomial whose
parameters are adjusted during the fitting procedure, and which
represents the real photospheric profile times the instrument
sensitivity; we denote this fitting model ``{\it U}'', which stands
for uncorrected (as the data have not been corrected prior to the fit
by a stellar model). In the second approach, we correct the data prior
to the fit by a stellar model and keep a free low-order polynomial
during the fit which models the inaccuracy of the stellar model and
the variation of the instrument sensitivity; we denote this approach
by ``{\it C}''. The calculated stellar model was shifted by
24.56~\kms\ in order to be consistent with the velocity of the
photospheric features clearly detected in N~{\sc v} and Si~{\sc iii}
(see Figure~\ref{STIS_3comp_Fig3}). These two approaches have been
followed in parallel to estimate the uncertainty in the modeling of
the stellar continuum on the final solution.

\begin{figure}
\begin{center}
\psfig{file=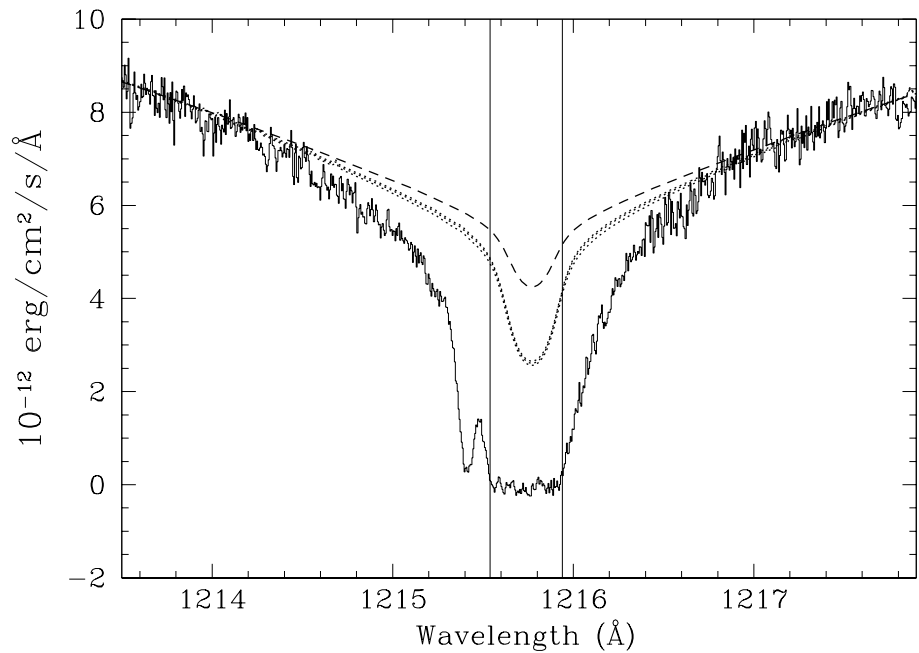,width=13cm}
\caption[]{The STIS data over the Lyman~$\alpha$ spectral region along
with the LTE (dashed line) and NLTE (dotted lines) photospheric
profiles. Three different NLTE profiles for atmospheric parameters in
the ranges $54000\,{\rm K}\,\leq T_{\rm eff}\leq 55000\,{\rm K}$ and
$7.5\leq \log g \leq 7.6$ are shown. The LTE calculation uses $T_{\rm
eff}=60880\,{\rm K}$ and $\log g = 7.59$. The radial velocity for the
photospheric profiles is set to $+24.56\,$\kms. The two thin vertical
lines delimit the wavelength range in which no information on the
stellar profile is contained in the data. The fitting procedure thus
tests only the difference between the wings of the models and not
their cores whose difference is more pronounced.}
\label{LyAandModel}
\end{center}
\end{figure}

We use two models to compute the photospheric Lyman\ $\alpha$ line
profile.  First, we use the program SYNSPEC (I.~Hubeny 2000, private
communication) in conjunction with the TLUSTY NLTE metal
line-blanketed model described above.  This version of SYNSPEC
contains the Lemke's Stark broadening tables for hydrogen, which were
computed by Lemke (1997) within the framework of Vidal, Cooper, \&
Smith (1973).  Second, we use Detlev Koester's LTE code and his best
fit parameters derived under the assumption of pure-H LTE ($T_{\rm
eff} = 60880\,{\rm K}$ and $\log g = 7.59$).  Figure~\ref{LyAandModel}
shows the NLTE and LTE Lyman\ $\alpha$ line profiles and illustrates
that NLTE effects are significant only in the core of the line, which
is formed high in the atmosphere where departures from LTE are
important (Wesemael {\it et al.} 1980).  However, as is shown in
Fig.~\ref{LyAandModel} the observations do not contain any information
in that central region of the photospheric line simply because it is
lost at the bottom of the saturated interstellar \lya\ line. Our
fitting procedure will however be able to test the slightly different
slopes in the wings of the models.

An additional correction made by a low-order polynomial is needed not
only to take into account uncertainties in the instrument sensitivity
but also because when one fits a section of the G191-B2B stellar
continuum devoid of stellar or interstellar absorption, it is
necessary to model the continuum with a polynomial of degree of order
1 to 3 for a spectral width of order 0.3\AA\ (see for instance
Fig.~\ref{STIS_3comp_Fig3}). For a spectral region of width
$\sim3$\AA\, as is the case for \lya, a 6$^{th}$ order polynomial
provides a satisfactory approximation to the stellar continuum.

The best-fit solutions obtained for 3 components on the line of sight
for all lines in the STIS domain including \lya\ and for various
models (prior correction or not of the \lya\ stellar continuum, type
of LSF) are summarized in Table~\ref{3compLyA}.  The zero point
wavelength of the Lyman $\alpha$ spectral domain was found to be
compatible with the other regions well within the $1\sigma$
uncertainty of 0.5\kms.

Concerning the fit with a free double gaussian (model 5 in
Table~\ref{3compLyA}), we found that the wings of the broader gaussian
(about 5 to 10 times larger than the narrow one) contribute in
amplitude (relatively to the amplitude of the narrow core gaussian) to
less than 1$\%$ in all but 5 spectral windows, less than 2$\%$ in 3 of
these 5 and to about 3$\%$ in the 2 remaining ones. Since this effect
is marginal, we kept only as comparison tests the use of both the
tabulated STIS LSF or the simple gaussian LSF.  All these effects were
included in the evaluation of the statistical errors.

It appears that different values of \nhi\ are obtained depending on
the model used; in particular the comparison of models 1 and 2 on the
one hand, and of models 3 and 4 on the other hand shows that the value
of \nhi\ derived depends whether the stellar continuum at \lya\ has
been corrected or not beforehand by a theoretical stellar profile. One
obtains $\log$\nhi=18.37 for an uncorrected profile, and
$\log$\nhi=18.32 for a corrected stellar profile. Since there is no
significant difference in \chid\ between these solutions, this
discrepancy must be interpreted as a systematic uncertainty tied to
the choice of modeling of the continuum. A $\Delta\chi^2$ analysis
around each of these solutions give a statistical error of $\pm0.01$
dex ($1\sigma$), so that the value of \nhi\ is dominated by the above
systematic error. Furthermore after correction of the stellar
continuum by an LTE calculation, which as shown in
Fig.~\ref{LyAandModel} represents an intermediate situation between
NLTE and non-corrected stellar profiles, the value of \nhi\ derived
was found to be an intermediate value between the above two.

However a further systematic error appears if additional weak hot
components are present on the line of sight, and its investigation is
the subject of the following Section.

\subsection{Systematic uncertainties and additional hot components}

\begin{figure}
\begin{center}
\psfig{file=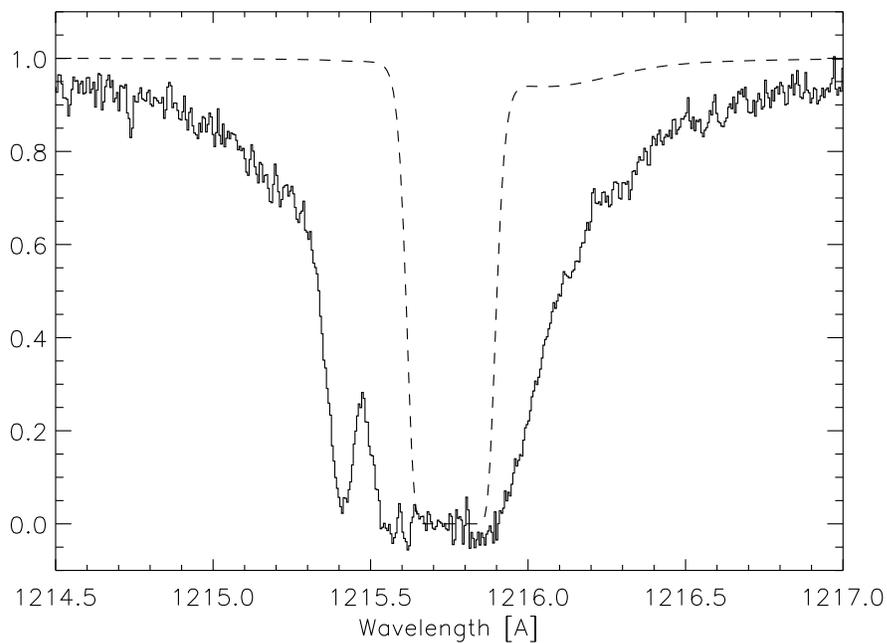,width=13cm}
\caption[]{The Lyman~$\alpha$ interstellar absorbtion observed with
STIS with the hydrogen wall model of Wood {\it et al.} (2000) for the
line of sight of G191-B2B overplotted in dashed line. The STIS data
has been normalized by the best-fit continuum.}
\label{Hwall}
\end{center}
\end{figure}

One cannot exclude the presence of weak H~{\sc i} components which
could perturb or bias the measurement of \nhi, but could not be
detected in any other species due to their weak column density.  Such
absorbers could arise as the signature of high velocity shocks (Cowie
{\it et al.} 1979), or as cloud interfaces with the hot gas within the
local ISM (Bertin {\it et al.} 1995) or as ``hydrogen walls'', i.e.,
the shock interaction between the solar wind (or stellar wind) and the
surrounding ISM (Linsky 1998). The latter possibility has been
modeled by Wood {\it et al.} (2000), and a prediction for the line of
sight toward G191-B2B is shown in Fig.~\ref{Hwall} overlaid on the
STIS data.  It shows that most of the expected absorption should take
place in the saturated core of the interstellar line, but some weak
absorption ($\sim5\%$) may be present, extending over several tenths
of an Angstrom on the red side of the line, due to the neutral
hydrogen atoms seen behind the shock in the downwind direction where
G191-B2B is located.

\begin{figure}
\begin{center}
\psfig{file=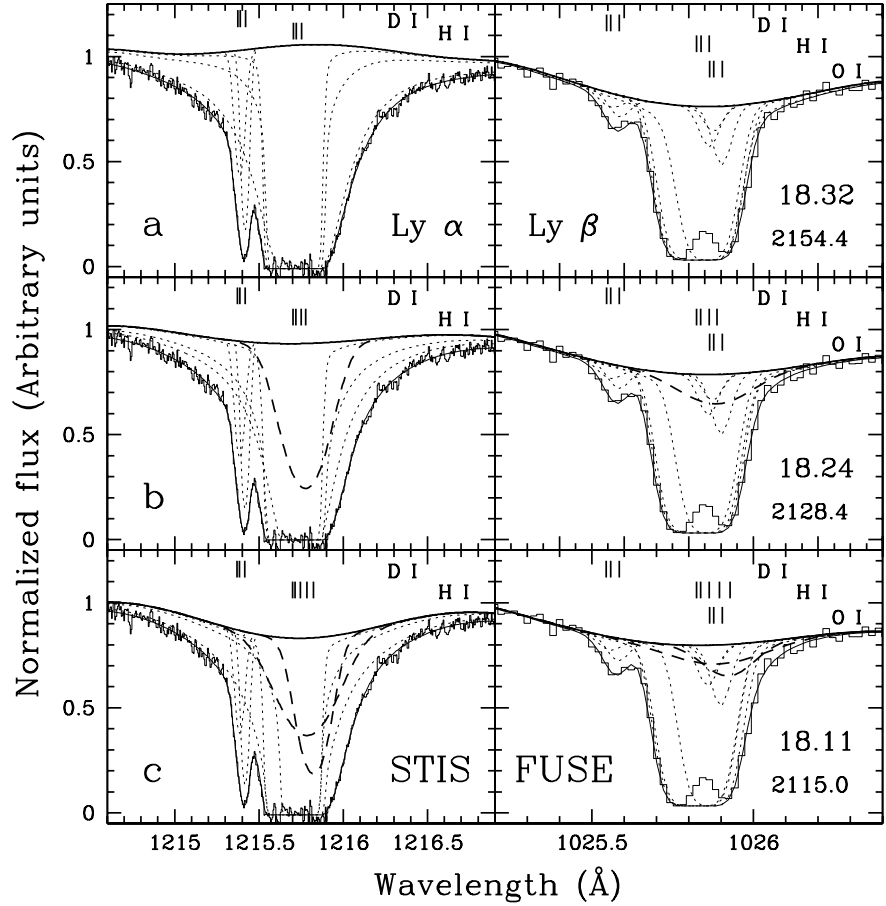,width=13cm}
\caption[]{Best-fit solutions for all lines in the STIS data plus four
\lyb\ from the {\it FUSE} HIRS dataset using three main interstellar
components plus zero, one and two extra H~{\sc i} components (upper,
middle, and lower panels respectively). The total absorption profile
is shown as the solid line, and the dotted and dashed lines show the
individual profiles (convolved with the LSF); the thick dashed lines
in particular show the contribution of the extra absorbers. The
solutions are labeled with the total rescaled \chid\ (for 1964, 1961,
and 1958 d.o.f. respectively for the upper, middle and lower panels)
and total H~{\sc i} column density (in log).}
\label{LyaLyb}
\end{center}
\end{figure}

  We have thus investigated the possible impact of additional weak hot
components on the determination of \nhi. In order to do so, we have
added one or two additional components in H~{\sc i} only, and performed
the fit of all lines as before with an NLTE correction of the \lya\
profile and using a free single Gaussian LSF (model 3). In order to
constrain the presence of these extra absorbers, we have added to the
fit the four \lyb\ lines observed through the {\it FUSE} HIRS
aperture.  For each new model characterized by the number of
additional components, we have measured \nhi\ using a $\Delta\chi^2$
analysis. The best fit solutions with zero, one and two extra
absorbers are shown for \lya\ and \lyb\ in Fig.~\ref{LyaLyb}.  We have
found that this introduction of additional components leads to a
substantial decrease of the total \nhi\ together with a significant
improvement of \chid.  In particular, we found that the best fit
solution for one extra component leads to $\Delta\chi^2=26$ (1961
d.o.f.) and $\log$\nhi=18.24, and two extra components lead to
$\Delta\chi^2=39.4$ (1958 d.o.f.) and $\log$\nhi=18.11. The \chid\
values quoted have been rescaled by a factor 1.53 corresponding to the
STIS data obtained in the previous section, and which remains close
(20\%) to the factor measured in the {\it FUSE} data range. Moreover,
if a $\Delta\chi^2$ analysis is performed for each model, the curve
obtained flattens for low values of \nhi\ down to $\log$\nhi=$18.0$;
in other words, the space of solutions becomes degenerate in the low
\nhi\ region, and consequently the value of \nhi\ measured from the
profile fitting of the \lya\ region is subject to a large systematic
uncertainty. This behaviour can be explained as follows. In the three
component best fit solution, most of the hydrogen is contained in the
LIC whose absorption makes the red wing of \lya. In the solutions with
extra absorbers, the combined profile of these hot components can
replace the damping wing contribution of the LIC, as shown in
Fig.~\ref{LyaLyb}.  

In order to scan the parameter space, the \nhi$_{\rm LIC}$ has been
forced to take various values, since the profile fitting code cannot
force the total H~{\sc i} column density to take a particular
value. When the LIC \nhi\ becomes negligible with regards to the total
\nhi, we force the next dominant component to take various values of
\nhi. In this way it is possible to scan the parameter space of the
total \nhi. One finds that once the LIC has become negligible,
component B2 makes most of the total \nhi, and that when the total
\nhi\ is taken below $18.0$, the \chid\ rapidly increases, which gives
the final error bar on the total \nhi\ estimate.

 At this stage, one should point out that these best-fit solutions are
quite difficult to find. The impact of additional weak absorbers on
the profile fitting of \nhi\ had already been studied by Vidal-Madjar
{\it et al.} (1998) but no significant effect had been found, as the
above solutions had eluded detection. The above solutions have
actually been obtained only in a late stage of the present work.  We
interpret this as evidence for the fact that the \chid\ surface
becomes complex when additional unconstrained absorbers are introduced
in the fit of the fully blended \lya\ profile and optimization is then
delicate.

  One should also remark that we have modeled these extra absorbers
using Voigt profiles, which implicitly assumes a Maxwellian
distribution for the velocities of the atoms. However if these
absorbers correspond to interstellar structures similar to hydrogen
walls, the latter assumption is incorrect and the overall profile
should be closer to that modeled by Wood {\it et al.} (2000) and shown
in Fig.~\ref{Hwall}. For this reason, the above modeling and solutions
should be interpreted as a first approximation. For this reason as
well, we do not discard the values for \nhi\ obtained in the previous
section without additional absorbers, even though the presence of
extra absorbers induce a significant gain in \chid. At the worst the
final value for \nhi\ will be more conservative. With these remarks in
mind, we conclude that the total hydrogen content toward G191-B2B
should be contained in the interval:

\begin{center}
$ 18.00 \leq \log N({\rm H}$~{\sc i}$) \leq 18.37 .$
\end{center}

In the absence of any indication on the distribution of the errors for
\nhi, we consider the above two extreme values as $2\sigma$ limits
since they correspond to rather extreme solutions, and quote the
following value for \nhi: $\log$\nhi$=18.18\pm0.09$ ($1\sigma$ error).
One should note that this effect of additional absorbers had not been
found in previous studies of the total hydrogen content toward
G191-B2B. Therefore it is important to remark that the above large
uncertainty must also affect the previous values of \nhi. In other
words, the above results supersedes any previous estimate of \nhi\
made toward G191-B2B using profile fitting of the \lya\ line. One may
also wonder whether extra absorbers may affect the determination of
\nhi\ toward other stars; this important question is the subject of a
forthcoming paper (Vidal-Madjar \& Ferlet 2002). On the other hand,
the above value for \nhi\ should be contrasted with the value derived
from EUVE observations and modeling of the atmosphere of G191-B2B
which gave $\log N$(H~{\sc i})=$18.315\pm0.013$ (Dupuis {\it et al.}
1995, 2$\sigma$ error), $\log N$(H~{\sc i})=$18.32$ (Lanz {\it et al.}
1996, no error bar quoted) and the recent detailed measurement $\log
N$(H~{\sc i})=$18.30\pm0.09$ (Barstow \& Hubeny 1998; this value also
agrees with the more recent work of Barstow, Hubeny \& Holberg
1999). These values would tend to indicate that the contamination of
the \lya\ profile by weak hot absorbers (if any) is not important, as
they agree with the values we obtained without including such extra
absorbers. However, as noted in Barstow \& Hubeny (1998), these fits
of the EUVE spectrum of G191-B2B typically lead to very large and
unexplained reduced $\chi^2$, implying that the overall fit is not yet
satisfactory in spite of the use of sophisticated WD atmosphere
models, and that some unknown effects have yet to be accounted for.

\section{Discussion and conclusions}

We have measured the total column densities of D~{\sc i}, N~{\sc i} and
O~{\sc i} toward G191-B2B using unsaturated absorption lines of these
elements in high quality {\it FUSE} spectra. After a careful
examination of the possible systematic uncertainties tied to the
choice of the stellar continuum and to the instrumental configuration,
we have derived the following column densities with $2\sigma$
uncertainties:

\begin{center}
$\log N({\rm D}$~{\sc i}$)_{\rm tot} = 13.40 \pm0.07,$ \\
$\log N({\rm O}$~{\sc i}$)_{\rm tot} = 14.86 \pm0.07,$ \\
$\log N({\rm N}$~{\sc i}$)_{\rm tot} = 13.87 \pm0.07.$ 
\end{center}

We have also analyzed new high signal-to-noise ratio high resolution
STIS observations of G191-B2B and provided concrete evidence for the
presence of at least three interstellar absorbing components on the
line of sight by analyzing the interstellar absorption lines of N~{\sc
i}, O~{\sc i}, Si~{\sc ii}, Si~{\sc iii}, S~{\sc iii} and Fe~{\sc ii}
present in the STIS bandpass. We have also measured the total hydrogen
column density on the line of sight using the velocity structure
derived from the above metals. We have performed an exhaustive study
of systematic effects on the value of \nhi. In particular we have
discovered a new major source of uncertainty on \nhi\ tied to the
possible presence of additional weak hot absorbers whose combined
absorption profile can contribute significantly to the wings of the
blended \lya\ profile. The column density of these absorbers is small
compared to the other main components, and they would not be detected
in any other species than H~{\sc i}, but their contribution to the
Lyman~$\alpha$ absorption profile can reduce significantly the total
H~{\sc i} column density measured from the profile fitting.  In order
to constrain their impact, we have analyzed simultaneously
Lyman~$\alpha$ and the higher order Lyman lines, and concluded that
the best value of \nhi\ toward G191-B2B is:

\begin{center}
$\log N({\rm H}$~{\sc i}$)_{\rm tot} = 18.18 \pm 0.18 \qquad
(2\sigma\,{\rm error})$
\end{center}

We emphasize that this uncertainty is a systematic uncertainty which
had gone unnoticed before. Therefore the above result supersedes
previous estimates of \nhi\ toward G191-B2B obtained from the profile
fitting of \lya.  A detailed analysis of this uncertainty and its
consequences on \nhi\ determinations toward other stars is discussed
in a companion paper (Vidal-Madjar \& Ferlet 2002). We thus derive
the following neutral abundance ratios toward G191-B2B, with $2\sigma$
uncertainties:

\begin{eqnarray}
({\rm D}/{\rm H})_{\rm tot}\, & = &\, 1.66^{+0.9}_{-0.6}\, 10^{-5}\nonumber \\
({\rm D}/{\rm O})_{\rm tot}\, & = &\, 3.49  \pm 0.78 \, 10^{-2}\nonumber \\
({\rm D}/{\rm N})_{\rm tot}\, & = &\, 3.41  \pm 0.76 \, 10^{-1}\nonumber \\
({\rm O}/{\rm H})_{\rm tot}\, & = &\, 4.79^{+2.5}_{-1.7}\, 10^{-4}\nonumber \\
({\rm N}/{\rm H})_{\rm tot}\, & = &\, 4.90^{+2.6}_{-1.8}\, 10^{-5}\nonumber \\
\end{eqnarray}

Most of the uncertainty in the above result results from the
systematic uncertainty on the \nhi\ determination.  This clearly shows
the importance of measuring accurate (D/O) and (D/N) ratios in the
interstellar medium instead of abundances relative to hydrogen, as
emphasized by Timmes {\it et al.} (1997). Interestingly if one uses
the recent measurement of \nhi\ from the modeling of the atmosphere of
G191-B2B and the fit of the EUVE spectrum, $\log N$(H~{\sc
i})$=18.30\pm0.09$ ($2\sigma$), one finds $({\rm D}/{\rm H})_{\rm
tot}=1.26_{-0.29}^{+0.36}\times 10^{-5}$ ($2\sigma$). At this stage,
however, due to the uncertainty inherent to the modeling of the white
dwarf atmosphere, it is probably more conservative to use the
interstellar determination for \nhi, and therefore the previous value
of the (D/H) ratio.


The above new value for the (D/H) ratio agree with the range of values
measured by Linsky (1998) toward a dozen stars of the LISM and with
the values previously derived toward G191-B2B. However the discrepancy
between previous estimates of the (D/H) ratio toward G191-B2B and the
LISM average D/H ratio has disappeared due to a revision of the
uncertainty on the estimation of the total H~{\sc i} content. A
detailed interpretation of this (D/H) value and of the accompanying
(D/O) and (D/N) ratios and their implications is provided in a
companion paper by Moos {\it et al.} (2002).

\acknowledgments 

  Financial support to U.S. participants has been provided
by NASA contract NAS5-32985. French participants are supported by
CNES.

\begin{deluxetable}{rccccc}
\tablecaption{Strongest N~{\sc i} and O~{\sc i} spectral lines in the {\it FUSE} 
domain.\label{FUSE-lines}}
\tablewidth{0pt}
\tablehead{
\colhead{Wavelength (\AA )} & 
\colhead{Element} & 
\colhead{$f$} & 
\colhead{$A$ (s$^{-1}$)} & 
\colhead{Comment} 
}
\startdata
1025.4440 & D~{\sc i}        & 0.264 $10^{-1}$  & 0.190 $10^{9}$&  \\ 
1025.4429 & D~{\sc i}          & 0.527 $10^{-1}$  & 0.190 $10^{9}$&  \\ 
 972.2725 & D~{\sc i}          & 0.967 $10^{-2}$  & 0.813 $10^{8}$&  \\ 
 972.2721 & D~{\sc i}          & 0.193 $10^{-1}$  & 0.813 $10^{8}$&  \\ 
 949.4848 & D~{\sc i}          & 0.465 $10^{-2}$  & 0.421 $10^{8}$& Weak  \\ 
 949.4846 & D~{\sc i}          & 0.929 $10^{-2}$  & 0.421 $10^{8}$& Weak  \\ 
\hline
1134.9803 & N~{\sc i}        & 0.435 $10^{-1}$ & 0.150 $10^{9}$&  \\
1134.4149 & N~{\sc i}        & 0.297 $10^{-1}$ & 0.154 $10^{9}$&  \\
1134.1653 & N~{\sc i}        & 0.152 $10^{-1}$ & 0.158 $10^{9}$&  \\
 954.1042 & N~{\sc i}        & 0.676 $10^{-2}$ & 0.330 $10^{8}$& Blend \\
 953.9699 & N~{\sc i}        & 0.348 $10^{-1}$ & 0.170 $10^{9}$&  \\
 953.6549 & N~{\sc i}        & 0.250 $10^{-1}$ & 0.183 $10^{9}$&  \\
 953.4152 & N~{\sc i}        & 0.132 $10^{-1}$ & 0.193 $10^{9}$&  \\
\hline
1039.2303 & O~{\sc i}        & 0.920 $10^{-2}$ & 0.947 $10^{8}$& \\
1026.4757 & O~{\sc i}        & 0.246 $10^{-2}$ & 0.111 $10^{8}$& Weak \\
1026.4744 & O~{\sc i}        & 0.187 $10^{-3}$ & 0.118 $10^{7}$& Weak \\
1025.7633 & O~{\sc i}        & 0.201 $10^{-3}$ & 0.212 $10^{7}$& Blend \\
1025.7626 & O~{\sc i}        & 0.302 $10^{-2}$ & 0.191 $10^{8}$& Blend \\
1025.7616 & O~{\sc i}        & 0.169 $10^{-1}$ & 0.765 $10^{8}$& Blend \\
 988.7734 & O~{\sc i}        & 0.465 $10^{-1}$ & 0.226 $10^{9}$& Strong \\
 988.6549 & O~{\sc i}        & 0.830 $10^{-2}$ & 0.566 $10^{8}$& Blend \\
 988.5778 & O~{\sc i}        & 0.553 $10^{-3}$ & 0.629 $10^{7}$& Weak \\
 976.4481 & O~{\sc i}        & 0.331 $10^{-2}$ & 0.386 $10^{8}$&  \\
 971.7382 & O~{\sc i}        & 0.116 $10^{-1}$ & 0.585 $10^{8}$& Strong \\
 971.7376 & O~{\sc i}        & 0.207 $10^{-2}$ & 0.146 $10^{8}$& Blend \\
 950.8846 & O~{\sc i}        & 0.158 $10^{-2}$ & 0.194 $10^{8}$&  \\
 948.6855 & O~{\sc i}        & 0.631 $10^{-2}$ & 0.100 $10^{9}$&  \\ 
 936.6295 & O~{\sc i}        & 0.365 $10^{-2}$ & 0.100 $10^{9}$&  \\
 929.5168 & O~{\sc i}        & 0.229 $10^{-2}$ & 0.100 $10^{9}$&  \\
 925.4460 & O~{\sc i}        & 0.354 $10^{-3}$ & 0.459 $10^{7}$& Weak \\
 924.9500 & O~{\sc i}       & 0.154 $10^{-2}$ & 0.100 $10^{9}$&  \\
 921.8570 & O~{\sc i}       & 0.100 $10^{-2}$ & 0.562 $10^{7}$& Weak \\
\enddata
\end{deluxetable}

\begin{deluxetable}{ccccc}
\tablecaption{{\it FUSE} observation log of G191-B2B.\label{FUSE-GBB}}
\tablewidth{0pt}
\tablehead{
\colhead{Dataset} & 
\colhead{Aperture} & 
\colhead{$T_{\rm exp}$ (ksec)} & 
\colhead{$N_{\rm exp}$} & 
\colhead{Date}  
}
\startdata
S3070101 & LWRS &  15.5 & 32 & 2000.01.14  \\
\hline
P1041202 & MDRS &  15.5 & 21 & 2000.01.13  \\
\hline
P1041201 & HIRS &  15.5 & 32 & 2000.11.06  \\
\enddata
\end{deluxetable}

\begin{deluxetable}{cccc}
\tablecaption{Published and {\it FUSE} O~{\sc i} and N~{\sc i} 
column densities.\label{OI_pub}}
\tablewidth{0pt}
\tablehead{
\colhead{Spectrograph $/$ } & 
\colhead{$\log N$(O~{\sc i})$_{\rm tot}$} & 
\colhead{$\log N$(N~{\sc i})$_{\rm tot}$} & 
\colhead{Reference}  \\
\colhead{Aperture} & 
\colhead{2$\sigma$ error} &
\colhead{2$\sigma$ error} &
\colhead{} 
}
\startdata
 HST GHRS-Ech. & 14.84$\pm0.04$ & 13.90$\pm0.02$ & 
Vidal--Madjar {\it et al.} 1998\tablenotemark{a}\\
\hline
 {\it FUSE} LWRS+MDRS & 14.79$\pm0.04$ & 13.82$\pm0.07$ & 
This work\tablenotemark{b}\\
\hline
 {\it FUSE} MDRS & 14.84$\pm0.08$ & 13.89$\pm0.06$ & 
This work\tablenotemark{b}\\
\hline
 {\it FUSE} HIRS & 14.88$\pm0.06$ & 13.84$\pm0.07$ & 
This work\tablenotemark{b}\\
\hline
 {\it FUSE}-All & 14.86$\pm0.07$ & 13.87$\pm0.07$ & 
This work\tablenotemark{b}\\
\enddata
\tablenotetext{a}{Three components were considered in the fit}
\tablenotetext{b}{One interstellar component was considered in the fit}
\end{deluxetable}

\begin{deluxetable}{cll}
\tablecaption{Published and {\it FUSE} D~{\sc i} column densities. 
\label{DI_pub}}
\tablewidth{0pt}
\tablehead{
\colhead{Spectrograph / } & 
\colhead{log $N$(D~{\sc i})$_{tot}$} & 
\colhead{Reference}  \\
\colhead{Spect./Aper.} & 
\colhead{2$\sigma$ error} & 
\colhead{/Model} 
}
\startdata
 HST GHRS-Ech. & 13.43$\pm0.02$ & 
Vidal--Madjar {\it et al.} 1998 \tablenotemark{a}\\
\hline
 HST STIS-Ech.\#1 & 13.55$^{+0.07}_{-0.08}$ &
Sahu {\it et al.} 1999 \tablenotemark{b}\\
\hline
 HST GHRS-Ech. & 13.40$\pm0.04$ &
Sahu {\it et al.} 1999 \tablenotemark{b}\\
\hline
 {\it FUSE} LWRS+MDRS & 13.41$^{+0.12}_{-0.07}$ & D~{\sc i}, no H~{\sc i},
this work \tablenotemark{c}\\
\hline
 {\it FUSE} MDRS & 13.41$\pm0.09$ &  D~{\sc i} \& H~{\sc i}, 
this work \tablenotemark{d}\\
\hline
 {\it FUSE} HIRS & 13.36$\pm0.08$ &  D~{\sc i} \& H~{\sc i} 
(Fig.~\ref{AAS_fig2}), 
this work \tablenotemark{d}\\
\hline
 {\it FUSE} HIRS & 13.38$\pm0.06$ & D~{\sc i} \& H~{\sc i}, double LSF, 
this work \tablenotemark{e}\\
\hline
 {\it FUSE} HIRS & 13.42$\pm0.08$ &  D~{\sc i}, no H~{\sc i}, 
this work \tablenotemark{c}\\
\hline
 {\it FUSE} All & 13.40$\pm0.07$ & this work \tablenotemark{f}\\
\enddata
\tablenotetext{a}{3 free absorbing components assumed in the profile
fitting, and stellar continuum modeled with a low order polynomial
free to vary during the fit}
\tablenotetext{b}{2 free absorbing components in the profile fitting,
and stellar continuum fixed and modeled by NLTE calculations}
\tablenotetext{c}{1 free absorbing component, stellar continuum
modeled by a freely varying low order polynomial, and D~{\sc i} fitted
alone without the H~{\sc i} line}
\tablenotetext{d}{1 free absorbing component, stellar continuum
modeled by a freely varying low order polynomial, and D~{\sc i} fitted
simultaneously with the H~{\sc i} line}
\tablenotetext{e}{1 free absorbing component, stellar continuum
modeled by a freely varying low order polynomial, and D~{\sc i} fitted
simultaneously with the H~{\sc i} line, LSF profile modeled with a 
double Gaussian}
\tablenotetext{f}{combination of FUSE models c, d and e}
\end{deluxetable}

\begin{deluxetable}{lcccc}
\tablecaption{Published H~{\sc i} column densities.\label{HI_pub}}
\tablewidth{0pt}
\tablehead{
\colhead{log $N$(H~{\sc i})$_{tot}$ ($2\sigma$)} &
\colhead{Spectrograph} &
\colhead{\# Comp.} &
\colhead{Continuum} &
\colhead{Reference}
}
\startdata
18.315$\pm0.013$ & EUVE & --- & --- & Dupuis {\it et al.} 1995 \\
\hline
18.32 -- & EUVE & --- & --- & Lanz {\it et al.} 1996 \\
\hline
18.36$\pm0.04$ & GHRS-G160M & 3 & free & Lemoine {\it et al.} 1996 \\
\hline
18.38$\pm0.02$ & GHRS-Ech. & 3 & free & Vidal--Madjar {\it et al.} 1998 \\
\hline
18.30$\pm0.09$ & EUVE & --- & --- & Barstow \& Hubeny 1998\\
\hline
18.31$\pm0.03$ & STIS-Ech. \#1 & 2 & fixed & Sahu {\it et al.} 1999 \\
\hline
18.34$\pm0.02$ & GHRS-Ech. & 2 & fixed & Sahu {\it et al.} 1999 \\
\enddata
\end{deluxetable}

\begin{deluxetable}{lllllccc}
\tablecaption{Spectral lines in the STIS domain used for the line of sight
structure study.\label{STISlines}}
\tablewidth{0pt}
\tablehead{
\colhead{Wavelength (\AA )} &
\colhead{Species} &
\colhead{{\it f}} &
\colhead{{\it A} (s$^{-1}$)} &
\colhead{Ech. Order}
}
\startdata
1334.5320 & C~{\sc ii}      & 0.128   & 0.288 $10^{9}$ & 316  \\
\hline
1200.7098 &  N~{\sc i}      & 0.0430   & 0.398 $10^{9}$ & 350, 351 \\
1200.2233 &  N~{\sc i}       & 0.0862   & 0.399 $10^{9}$ & 350, 351 \\
1199.5496 &  N~{\sc i}      & 0.130   & 0.401 $10^{9}$ & 351 \\
\hline
1242.8040 &  N~{\sc v}      & 0.500   & 0.336 $10^{9}$ & 339 \\
\hline
1302.1685 &  O~{\sc i}      & 0.0519   & 0.340 $10^{9}$ & 323 \\
\hline
1304.3702 &  Si~{\sc ii}     & 0.0917   & 0.107 $10^{10}$ & 323 \\
1193.2897 &  Si~{\sc ii}     & 0.585   & 0.409 $10^{10}$ & 352, 353 \\
1190.4158 &  Si~{\sc ii}     & 0.293   & 0.410 $10^{10}$ & 353, 354 \\
\hline
1206.5000 &  Si~{\sc iii}    & 1.67   & 0.255 $10^{10}$ & 349\\
\hline
1259.5190 &  S~{\sc ii}     & 0.0166   & 0.465 $10^{8}$ & 334 \\
\hline
1190.2030 &  S~{\sc iii}    & 0.0231   & 0.651 $10^{8}$ & 353, 354\\
\hline
2382.7651 &  Fe~{\sc ii}     & 0.320   & 0.313 $10^{9}$ & 324 \\
2586.6499 &  Fe~{\sc ii}     & 0.0691  & 0.272 $10^{9}$ & 298, 299 \\
2600.1729 &  Fe~{\sc ii}     & 0.239   & 0.270 $10^{9}$ & 297 \\
\enddata
\end{deluxetable}

\begin{deluxetable}{lllcllc}
\tablecaption{\chid\ comparison for the two and three component
solutions using either the tabulated STIS LSF or freely varying single
gaussian LSF. \label{chi2}}
\tablewidth{0pt}
\tablehead{
\colhead{Species} & 
\colhead{$\chi^2$ / d.o.f.}\tablenotemark{a} & 
\colhead{$\chi^2$ / d.o.f.}\tablenotemark{a} &  
\colhead{$F-$test prob.}\tablenotemark{b} &  
\colhead{$\chi^2$ / d.o.f.}\tablenotemark{a} & 
\colhead{$\chi^2$ / d.o.f.}\tablenotemark{a} &  
\colhead{$F-$test prob.}\tablenotemark{b} \\  
\colhead{} &
\colhead{2 comp.} & 
\colhead{3 comp.} &  
\colhead{2 {\it vs} 3 comp.} &  
\colhead{2 comp.} &  
\colhead{3 comp.} &  
\colhead{2 {\it vs} 3 comp.} \\  
\colhead{} & 
\colhead{STIS LSF} &  
\colhead{STIS LSF} &  
\colhead{} &  
\colhead{free Gaussian} &  
\colhead{free Gaussian} &  
\colhead{} \\  
}
\startdata
N~{\sc i}    & 395.0/314  & 382.3/311 & 1.7\% & 323.2/309 & 309.1/306 & 0.3\% \\
O~{\sc i}    & 214.3/61   & 193.1/58 &  11\% & 113.9/60 & 101.4/57 & 8.3\% \\
Si~{\sc ii}  & 772.0/468  & 651.4/465 & $<0.01$\% & 678.6/463 &
553.6/460 &  $<0.01$\% \\
Si~{\sc iii}\tablenotemark{c} & 210.6/152  & 192.8/149 & 0.4\% &
185.2/150 & 170.8/147 & 0.75\% \\
Fe~{\sc ii}  & 277.6/142  & 141.8/139 & $<0.01$\%  & 227.9/139 & 
123.3/136 & $<0.01$\% \\
\hline
All\tablenotemark{d}     & 2677.8/1270 & 1908.8/1259 & 
$<0.01$\% & 2076.7/1250 & 1442.5/1239 & $<0.01$\% \\
\enddata
\tablenotetext{a}{the $\chi^2$ has been rescaled to compensate for the 
inaccuracy of the noise array (see text)}
\tablenotetext{b}{The $F-$test gives the
probability that a third absorbing component is not required by the model} 
\tablenotetext{c}{Fitted simultaneously with the N~{\sc v} 
line to control the photospheric
Si~{\sc iii} line}
\tablenotetext{d}{Includes  C~{\sc ii}, S~{\sc ii} and S~{\sc iii}}
\end{deluxetable}

\begin{deluxetable}{cccccc}
\tablecaption{Three component best-fit solutions for all lines in the
STIS region including \lya. \label{3compLyA}}
\tablewidth{0pt}
\tablehead{
\colhead{Fit} & 
\colhead{$v_{\rm B1}$} & 
\colhead{$v_{\rm B2}$} & 
\colhead{$v_{\rm LIC}$} & 
\colhead{log$N$(H~{\sc i})$_{\rm tot}$} &  
\colhead{$\chi^2$} \\
\colhead{model} &
\colhead{(\kms)} &
\colhead{(\kms)} &
\colhead{(\kms)} &
\colhead{} &  
\colhead{1753 d.o.f.}  
}
\startdata
 1\tablenotemark{a,c} & 7.7 &  11.2 &  19.4 &  18.32 & 2469.9 \\
 2\tablenotemark{b,c} & 7.7 &  11.3 &  19.4 &  18.37 & 2470.1 \\
\hline
 3\tablenotemark{a,d} & 7.7 &  11.7 &  19.4 &  18.32 & 2006.8 \\
 4\tablenotemark{b,d} & 7.7 &  11.6 &  19.4 &  18.37 & 2001.8 \\
\hline
 5\tablenotemark{a,e} & 7.7 &  11.6 &  19.4 &  18.33 & 1987.3 \\
\enddata
\tablenotetext{a}{data normalized beforehand by NLTE stellar continuum, and 
stellar continuum residuals modeled during the fit by a 6$^{\rm th}$ order 
polynomial}
\tablenotetext{b}{unnormalized data, with stellar continuum modeled
during the fit by a 6$^{\rm th}$ order polynomial}
\tablenotetext{c}{LSF corresponds to the tabulated STIS LSF}
\tablenotetext{d}{LSF profile modeled with a single Gaussian (free
to vary during the fit)}
\tablenotetext{e}{LSF profile modeled with a double Gaussian (free to
vary during the fit)} 
\end{deluxetable}


\begin{thebibliography}{}
 

\bibitem{aa1992} Allen, M.M., Jenkins, E.B., \& Snow, T.P., 1992, ApJS,
83, 261

\bibitem{at} Audouze, J., Tinsley, B. M., 1976, ARA\&A, 14, 43

\bibitem[]{b98} Barstow, M.A., Hubeny, I., \& Holberg, J. B., 1998,
MNRAS, 299, 520

\bibitem[]{bhk95} Barstow, M. A., Holberg, J. B., \& Koester, D., 1995,
MNRAS, 274, L31

\bibitem[]{bh98} Barstow, M. A., Hubeny, I., 1998, MNRAS, 299, 379

\bibitem[]{bhh99} Barstow, M. A., Hubeny, I., Holberg, J. B., 1999,
MNRAS 307, 884

\bibitem[]{sb01} Barstow, M. A., {\it et al.}\ 2001, in {\it The
12$^{\rm th}$ European Workshop on White Dwarfs}, ASP Conf. Series, in
press
   

\bibitem[]{b95} Bertin, P., {\it et al.}, 1995, A\&A, 302, 889

\bibitem{bs} Boesgaard, A. M.,  Steigman, G., 1985, ARA\&A, 23, 319 

\bibitem{b01} Burles, S., 2001, in {\em Gaseous matter in galaxies and
intergalactic space - XVII$^{th}$ IAP Colloquium} (Paris, June 19-23
2001), eds. R.~Ferlet {\it et al.} (Frontier Group), in press.

\bibitem{cv} Cass\'e, M., \& Vangioni-Flam, E., 1998, in {\em Structure
and Evolution of the Intergalactic Medium from QSO Absorption Line
Systems}, eds. P. Petitjean and S. Charlot (IAP Conference series),
331

\bibitem{co} Cowie, L., {\it et al.}, 1979, ApJ, 229, L81


\bibitem{da5} Dupuis, J., {\it et al.}, 1995, ApJ, 455, 574

\bibitem{da8} Dupuis, J., {\it et al.}, 1998, ApJ, 500, L45

\bibitem{fa1980} Ferlet, R., {\it et al.}, 1980, ApJ, 242, 576

\bibitem{fa1} Friedman, S.D., {\it et al.}, 2002, accepted for publication in ApJS

\bibitem{go} Gautier, D., Owen, T., 1983, Nature, 302, 215

\bibitem{gr} Geiss, J., Reeves, H., 1972, A\&A, 18, 126

\bibitem{glvm} Gry, C., Lamers, H. J. G. L. M., Vidal--Madjar, A.,
1984, A\&A, 137, 29

\bibitem{h99} H\'ebrard, G., Mallouris, C., Ferlet, R., Koester, D., Lemoine,
M., Vidal--Madjar, A., \& York, D. G., 1999, A\&A, 350, 643

\bibitem{ha1} H\'ebrard, G., {\it et al.}, 2002, accepted for publication in ApJS

\bibitem[]{hs00} Howk, J. C., Sembach, K. R., 2000, AJ, 119, 2481

\bibitem[]{hl95} Hubeny, I., Lanz, T., 1995, ApJ, 439, 875

\bibitem[]{j99} Jenkins, E. B., Tripp, T. M., Wo\'zniak, P. R., Sofia,
U. J., Sonneborn, G., 1999, ApJ, 520, 182

\bibitem{ka1} Kruk, J. W., {\it et al.}, 2002, accepted for publication in ApJS

\bibitem{lsb96} Landsman, W., Sofia, U. J., Bergeron, P., 1996, in 
{\em Science with the Hubble Space Telescope - II}, STScI, 454

\bibitem[]{lb96} Lanz, T., Barstow, M. A., Hubeny, I., Holberg, J.B.,
 1996, ApJ, 473, 1089 

\bibitem{lvy} Laurent, C., Vidal--Madjar, A., York, D.G.,
1979, ApJ, 229, 923

\bibitem[]{l01} Lehner, N., {\it et al.}, 2002, accepted for publication in ApJS

\bibitem[]{l97} Lemke, M., 1997, A\&AS, 122, 285
   
\bibitem[]{l96} Lemoine, M., Vidal--Madjar, A., Bertin, P., Ferlet,
R., Gry, C., \& Lallement, R., 1996, A\&A, 308, 601

\bibitem{la9} Lemoine, M., {\it et al.}, 1999, New Astronomy, 4, 231


\bibitem{l98} Linsky, J. L., 1998, Space Science Reviews, 84, 285



\bibitem[]{m00} Moos, H.W., {\it et al.}, 2000, ApJ, 538, L1

\bibitem{ma1} Moos, H.W., {\it et al.}, 2002, accepted for publication in ApJS

\bibitem{pa} Pagel, B., {\it et al.}, 1992, MNRAS, 255, 325

\bibitem{p} Prantzos, N., 1996, A\&A, 310, 106

\bibitem{ry} Rogerson, J., York, D. G., 1973, ApJ, 186, L95

\bibitem[]{s00} Sahnow, D.J., {\it et al.},  2000, ApJ, 538, L7

\bibitem[]{s99} Sahu, M. S., Landsman, W., Bruhweiler, F. C., Gull,
T. R., Bowers, C. A., Lindler, D., Feggans, K., Barstow, M. A., Hubeny,
I., Holberg, J. B., 1999, ApJ, 523, L159

\bibitem{sa0} Sahu, M. S., 2000, in {\em The Light Elements and
Their Evolution}, eds. L.~da~Silva, M.~Spite,
 J.~R.~de~Medeiros (ASP Conference Series), 161

\bibitem{sc} Scully, S. T., {\it et al.}, 1997, ApJ, 476, 521

\bibitem[]{stfjsvm00} Sonneborn, G., Tripp, T. M., Ferlet, R., Jenkins,
E. B., Sofia, U.  J., Vidal--Madjar, A.,  Wo\'zniak, P. R., 2000, ApJ,
545, 277

\bibitem{sa1} Sonneborn, G., {\it et al.}, 2002, ApJS, submitted

\bibitem{ttly97} Timmes, F. X., Truran, J. W., Lauroesch, J. T., York,
D. G., 1997, ApJ, 476, 464

\bibitem{tosi} Tosi, M., Steigman, G., Matteucci,F., Chiappini,
C., 1998, ApJ 498, 226


\bibitem{vc4} Vangioni--Flam, E., \& Cass\'e, M., 1994, ApJ, 427, 618

\bibitem{vc5} Vangioni--Flam, E., Olive, K., \& Prantzos, N., 1995,
ApJ, 441, 471

\bibitem[]{v00} Vennes, S., Polomski, E.\ F., Lanz, T., Thorstensen,
J.\ R., Chayer, P., Gull, T.\ R., 2000, ApJ, 544, 423

\bibitem[]{v96} Vennes, S., Chayer, P., Hurwitz, M., Bowyer, S., 1996,
ApJ, 468, 898
 
\bibitem[]{v73} Vidal, C.\ R., Cooper, J., Smith, E.\ W., 1973, ApJS,
25, 37

\bibitem{va84} Vidal--Madjar, A., Gry, C., 1984, A\&A, 138, 285 

\bibitem{va77} Vidal--Madjar, A., {\it et al.}, 1977, ApJ, 211, 91 

\bibitem[]{v98} Vidal--Madjar, A., Lemoine, M., Ferlet, R., H\'ebrard,
G., Koester, D., Audouze, J., Cass\'e, M., Vangioni-Flam, E., Webb,
J. K., 1998, A\&A, 338, 694

\bibitem{v} Vidal--Madjar, A., 2000, in {\em The Light Elements and
Their Evolution}, eds. L.~da~Silva, M.~Spite, \& J.~R.~de~Medeiros
(ASP Conference Series), 151

\bibitem{v01} Vidal--Madjar, A., 2001, in {\em Cosmic Evolution},
eds. E. Vangioni-Flam, R. Ferlet, M. Lemoine (World Scientific), p.49

\bibitem{vf01} Vidal--Madjar, A., Ferlet, R., 2002, in preparation

\bibitem{y1983} York, D. G., 1983, ApJ, 264, 172

\bibitem{yr} York, D.G., Rogerson, J., 1976, ApJ, 203, 378


\bibitem[]{w80} Wesemael, F., Auer, L.\ H., Van Horn, H.\ M.,
Savedoff, M.\ P., 1980, ApJS, 43, 159
   

\bibitem{la1} Wood, B. E., Muller, H--R., Zank, G. P., 2000, ApJ, 542, 493

\bibitem{w01} Wood, B. E., {\it et al.}, 2002, accepted for publication in ApJS

\end{thebibliography}
\end{document}